\documentclass[%
 reprint,
superscriptaddress,
nofootinbib,
 amsmath,amssymb,
 aps,
prd,
]{revtex4-2}
\usepackage[utf8]{inputenc}
\usepackage[margin=1in]{geometry}
\usepackage{physics, bbold, graphicx, amsmath, amsthm, amssymb, amsfonts, siunitx, enumerate,listings, multirow, float, mathtools, subfigure}
\usepackage[ruled,lined]{algorithm2e}
\usepackage{dcolumn}
\usepackage{babel}
 
\usepackage{orcidlink}
\newcommand{\snowglobes}{SNOwGLoBES}
\newcommand{\larsoft}{LArSoft}
\newcommand{\arforty}{{}^{40}\mathrm{Ar}}

\begin{document}

\preprint{APS/123-QED}
\title{Supernova Pointing Capabilities of DUNE}

%

\newcommand{\Abilene}{Abilene Christian University, Abilene, TX 79601, USA}
\newcommand{\Albanysuny}{University of Albany, SUNY, Albany, NY 12222, USA}
\newcommand{\Amsterdam}{University of Amsterdam, NL-1098 XG Amsterdam, The Netherlands}
\newcommand{\Antalya}{Antalya Bilim University, 07190 D{\"o}{\c{s}}emealt{\i}/Antalya, Turkey}
\newcommand{\Antananarivo}{University of Antananarivo, Antananarivo 101, Madagascar}
\newcommand{\Antioquia}{University of Antioquia, Medell{\'\i}n, Colombia}
\newcommand{\AntonioNarino}{Universidad Antonio Nari{\~n}o, Bogot{\'a}, Colombia}
\newcommand{\Argonne}{Argonne National Laboratory, Argonne, IL 60439, USA}
\newcommand{\Arizona}{University of Arizona, Tucson, AZ 85721, USA}
\newcommand{\Asuncion}{Universidad Nacional de Asunci{\'o}n, San Lorenzo, Paraguay}
\newcommand{\Athens}{University of Athens, Zografou GR 157 84, Greece}
\newcommand{\Atlantico}{Universidad del Atl{\'a}ntico, Barranquilla, Atl{\'a}ntico, Colombia}
\newcommand{\Augustana}{Augustana University, Sioux Falls, SD 57197, USA}
\newcommand{\Bern}{University of Bern, CH-3012 Bern, Switzerland}
\newcommand{\Beykent}{Beykent University, Istanbul, Turkey}
\newcommand{\Birmingham}{University of Birmingham, Birmingham B15 2TT, United Kingdom}
\newcommand{\BolognaUniversity}{Universit{\`a} di Bologna, 40127 Bologna, Italy}
\newcommand{\Boston}{Boston University, Boston, MA 02215, USA}
\newcommand{\Bristol}{University of Bristol, Bristol BS8 1TL, United Kingdom}
\newcommand{\Brookhaven}{Brookhaven National Laboratory, Upton, NY 11973, USA}
\newcommand{\Bucharest}{University of Bucharest, Bucharest, Romania}
\newcommand{\CalBerkeley}{University of California Berkeley, Berkeley, CA 94720, USA}
\newcommand{\CalDavis}{University of California Davis, Davis, CA 95616, USA}
\newcommand{\CalIrvine}{University of California Irvine, Irvine, CA 92697, USA}
\newcommand{\CalLosangeles}{University of California Los Angeles, Los Angeles, CA 90095, USA}
\newcommand{\CalRiverside}{University of California Riverside, Riverside CA 92521, USA}
\newcommand{\CalSantabarbara}{University of California Santa Barbara, Santa Barbara, California 93106 USA}
\newcommand{\Caltech}{California Institute of Technology, Pasadena, CA 91125, USA}
\newcommand{\Cambridge}{University of Cambridge, Cambridge CB3 0HE, United Kingdom}
\newcommand{\Campinas}{Universidade Estadual de Campinas, Campinas - SP, 13083-970, Brazil}
\newcommand{\CataniaUniversitadi}{Universit{\`a} di Catania, 2 - 95131 Catania, Italy}
\newcommand{\Catolica}{Universidad Cat{\'o}lica del Norte, Antofagasta, Chile}
\newcommand{\CBPF}{Centro Brasileiro de Pesquisas F\'isicas, Rio de Janeiro, RJ 22290-180, Brazil}
\newcommand{\CEASaclay}{IRFU, CEA, Universit{\'e} Paris-Saclay, F-91191 Gif-sur-Yvette, France}
\newcommand{\CERN}{CERN, The European Organization for Nuclear Research, 1211 Meyrin, Switzerland}
\newcommand{\Charles}{Institute of Particle and Nuclear Physics of the Faculty of Mathematics and Physics of the Charles University, 180 00 Prague 8, Czech Republic }
\newcommand{\Chicago}{University of Chicago, Chicago, IL 60637, USA}
\newcommand{\ChungAng}{Chung-Ang University, Seoul 06974, South Korea}
\newcommand{\CIEMAT}{CIEMAT, Centro de Investigaciones Energ{\'e}ticas, Medioambientales y Tecnol{\'o}gicas, E-28040 Madrid, Spain}
\newcommand{\Cincinnati}{University of Cincinnati, Cincinnati, OH 45221, USA}
\newcommand{\Cinvestav}{Centro de Investigaci{\'o}n y de Estudios Avanzados del Instituto Polit{\'e}cnico Nacional (Cinvestav), Mexico City, Mexico}
\newcommand{\Colima}{Universidad de Colima, Colima, Mexico}
\newcommand{\ColoradoBoulder}{University of Colorado Boulder, Boulder, CO 80309, USA}
\newcommand{\ColoradoState}{Colorado State University, Fort Collins, CO 80523, USA}
\newcommand{\Columbia}{Columbia University, New York, NY 10027, USA}
\newcommand{\conida}{Comisi{\'o}n Nacional de Investigaci{\'o}n y Desarrollo Aeroespacial, Lima, Peru}
\newcommand{\Cti}{Centro de Tecnologia da Informacao Renato Archer, Amarais - Campinas, SP - CEP 13069-901}
\newcommand{\CUSB}{Central University of South Bihar, Gaya, 824236, India }
\newcommand{\CzechAcademyofSciences}{Institute of Physics, Czech Academy of Sciences, 182 00 Prague 8, Czech Republic}
\newcommand{\CzechTechnical}{Czech Technical University, 115 19 Prague 1, Czech Republic}
\newcommand{\DannecyleVieux}{Laboratoire d{\textquoteright}Annecy de Physique des Particules, Universit{\'e} Savoie Mont Blanc, CNRS, LAPP-IN2P3, 74000 Annecy, France}
\newcommand{\Daresbury}{Daresbury Laboratory, Cheshire WA4 4AD, United Kingdom}
\newcommand{\Dordt}{Dordt University, Sioux Center, IA 51250, USA}
\newcommand{\Drexel}{Drexel University, Philadelphia, PA 19104, USA}
\newcommand{\Duke}{Duke University, Durham, NC 27708, USA}
\newcommand{\Durham}{Durham University, Durham DH1 3LE, United Kingdom}
\newcommand{\Edinburgh}{University of Edinburgh, Edinburgh EH8 9YL, United Kingdom}
\newcommand{\EIA}{Universidad EIA, Envigado, Antioquia, Colombia}
\newcommand{\erciyes}{Erciyes University, Kayseri, Turkey}
\newcommand{\Eotvos}{E{\"o}tv{\"o}s Lor{\'a}nd University, 1053 Budapest, Hungary}
\newcommand{\FCULport}{Faculdade de Ci{\^e}ncias da Universidade de Lisboa - FCUL, 1749-016 Lisboa, Portugal}
\newcommand{\FederaldeAlfenas}{Universidade Federal de Alfenas, Po{\c{c}}os de Caldas - MG, 37715-400, Brazil}
\newcommand{\FederaldeGoias}{Universidade Federal de Goias, Goiania, GO 74690-900, Brazil}
\newcommand{\FederaldoABC}{Universidade Federal do ABC, Santo Andr{\'e} - SP, 09210-580, Brazil}
\newcommand{\FederaldoRio}{Universidade Federal do Rio de Janeiro,  Rio de Janeiro - RJ, 21941-901, Brazil}
\newcommand{\Fermi}{Fermi National Accelerator Laboratory, Batavia, IL 60510, USA}
\newcommand{\Ferrarauniv}{University of Ferrara, Ferrara, Italy}
\newcommand{\Florida}{University of Florida, Gainesville, FL 32611-8440, USA}
\newcommand{\Floridastate}{Florida State University, Tallahassee, FL, 32306 USA}
\newcommand{\Fluminense}{Fluminense Federal University, 9 Icara{\'\i} Niter{\'o}i - RJ, 24220-900, Brazil }
\newcommand{\Genova}{Universit{\`a} degli Studi di Genova, Genova, Italy}
\newcommand{\Georgian}{Georgian Technical University, Tbilisi, Georgia}
\newcommand{\Granada}{University of Granada {\&} CAFPE, 18002 Granada, Spain}
\newcommand{\GranSasso}{Gran Sasso Science Institute, L'Aquila, Italy}
\newcommand{\GranSassoLab}{Laboratori Nazionali del Gran Sasso, L'Aquila AQ, Italy}
\newcommand{\Grenoble}{University Grenoble Alpes, CNRS, Grenoble INP, LPSC-IN2P3, 38000 Grenoble, France}
\newcommand{\Guanajuato}{Universidad de Guanajuato, Guanajuato, C.P. 37000, Mexico}
\newcommand{\Harish}{Harish-Chandra Research Institute, Jhunsi, Allahabad 211 019, India}
\newcommand{\Hawaii}{University of Hawaii, Honolulu, HI 96822, USA}
\newcommand{\hkust}{Hong Kong University of Science and Technology, Kowloon, Hong Kong, China}
\newcommand{\Houston}{University of Houston, Houston, TX 77204, USA}
\newcommand{\Hyderabad}{University of  Hyderabad, Gachibowli, Hyderabad - 500 046, India}
\newcommand{\Idaho}{Idaho State University, Pocatello, ID 83209, USA}
\newcommand{\IFIC}{Instituto de F{\'\i}sica Corpuscular, CSIC and Universitat de Val{\`e}ncia, 46980 Paterna, Valencia, Spain}
\newcommand{\IGFAE}{Instituto Galego de F{\'\i}sica de Altas Enerx{\'\i}as, University of Santiago de Compostela, Santiago de Compostela, 15782, Spain}
\newcommand{\Iitk}{Indian Institute of Technology Kanpur, Uttar Pradesh 208016, India}
\newcommand{\Illinoisinstitute}{Illinois Institute of Technology, Chicago, IL 60616, USA}
\newcommand{\Imperial}{Imperial College of Science Technology and Medicine, London SW7 2BZ, United Kingdom}
\newcommand{\IndGuwahati}{Indian Institute of Technology Guwahati, Guwahati, 781 039, India}
\newcommand{\IndHyderabad}{Indian Institute of Technology Hyderabad, Hyderabad, 502285, India}
\newcommand{\Indiana}{Indiana University, Bloomington, IN 47405, USA}
\newcommand{\INFNBologna}{Istituto Nazionale di Fisica Nucleare Sezione di Bologna, 40127 Bologna BO, Italy}
\newcommand{\INFNCatania}{Istituto Nazionale di Fisica Nucleare Sezione di Catania, I-95123 Catania, Italy}
\newcommand{\INFNFerrara}{Istituto Nazionale di Fisica Nucleare Sezione di Ferrara, I-44122 Ferrara, Italy}
\newcommand{\INFNFrascati}{Istituto Nazionale di Fisica Nucleare Laboratori Nazionali di Frascati, Frascati, Roma, Italy}
\newcommand{\INFNGenova}{Istituto Nazionale di Fisica Nucleare Sezione di Genova, 16146 Genova GE, Italy}
\newcommand{\INFNLecce}{Istituto Nazionale di Fisica Nucleare Sezione di Lecce, 73100 - Lecce, Italy}
\newcommand{\INFNMilanBicocca}{Istituto Nazionale di Fisica Nucleare Sezione di Milano Bicocca, 3 - I-20126 Milano, Italy}
\newcommand{\INFNMilano}{Istituto Nazionale di Fisica Nucleare Sezione di Milano, 20133 Milano, Italy}
\newcommand{\INFNNapoli}{Istituto Nazionale di Fisica Nucleare Sezione di Napoli, I-80126 Napoli, Italy}
\newcommand{\INFNPadova}{Istituto Nazionale di Fisica Nucleare Sezione di Padova, 35131 Padova, Italy}
\newcommand{\INFNPavia}{Istituto Nazionale di Fisica Nucleare Sezione di Pavia,  I-27100 Pavia, Italy}
\newcommand{\INFNPisa}{Istituto Nazionale di Fisica Nucleare Laboratori Nazionali di Pisa, Pisa PI, Italy}
\newcommand{\INFNRoma}{Istituto Nazionale di Fisica Nucleare Sezione di Roma, 00185 Roma RM, Italy}
\newcommand{\INFNSud}{Istituto Nazionale di Fisica Nucleare Laboratori Nazionali del Sud, 95123 Catania, Italy}
\newcommand{\Ingenieria}{Universidad Nacional de Ingenier{\'\i}a, Lima 25, Per{\'u}}
\newcommand{\Insubria }{University of Insubria, Via Ravasi, 2, 21100 Varese VA, Italy}
\newcommand{\Iowa}{University of Iowa, Iowa City, IA 52242, USA}
\newcommand{\IowaState}{Iowa State University, Ames, Iowa 50011, USA}
\newcommand{\IPLyon}{Institut de Physique des 2 Infinis de Lyon, 69622 Villeurbanne, France}
\newcommand{\IPM}{Institute for Research in Fundamental Sciences, Tehran, Iran}
\newcommand{\ISTlisboa}{Instituto Superior T{\'e}cnico - IST, Universidade de Lisboa, Portugal}
\newcommand{\Ita}{Instituto Tecnol{\'o}gico de Aeron{\'a}utica, Sao Jose dos Campos, Brazil}
\newcommand{\Iwate}{Iwate University, Morioka, Iwate 020-8551, Japan}
\newcommand{\Jacksonstate}{Jackson State University, Jackson, MS 39217, USA}
\newcommand{\Jawaharlal}{Jawaharlal Nehru University, New Delhi 110067, India}
\newcommand{\Jeonbuk}{Jeonbuk National University, Jeonrabuk-do 54896, South Korea}
\newcommand{\Jyvaskyla}{Jyv{\"a}skyl{\"a} University, FI-40014 Jyv{\"a}skyl{\"a}, Finland}
\newcommand{\Kansasstate}{Kansas State University, Manhattan, KS 66506, USA}
\newcommand{\Kavli}{Kavli Institute for the Physics and Mathematics of the Universe, Kashiwa, Chiba 277-8583, Japan}
\newcommand{\KEK}{High Energy Accelerator Research Organization (KEK), Ibaraki, 305-0801, Japan}
\newcommand{\KISTI}{Korea Institute of Science and Technology Information, Daejeon, 34141, South Korea}
\newcommand{\Kure}{National Institute of Technology, Kure College, Hiroshima, 737-8506, Japan}
\newcommand{\Kyiv}{Taras Shevchenko National University of Kyiv, 01601 Kyiv, Ukraine}
\newcommand{\Lancaster}{Lancaster University, Lancaster LA1 4YB, United Kingdom}
\newcommand{\LawrenceBerkeley}{Lawrence Berkeley National Laboratory, Berkeley, CA 94720, USA}
\newcommand{\LIP}{Laborat{\'o}rio de Instrumenta{\c{c}}{\~a}o e F{\'\i}sica Experimental de Part{\'\i}culas, 1649-003 Lisboa and 3004-516 Coimbra, Portugal}
\newcommand{\Liverpool}{University of Liverpool, L69 7ZE, Liverpool, United Kingdom}
\newcommand{\LosAlmos}{Los Alamos National Laboratory, Los Alamos, NM 87545, USA}
\newcommand{\Louisanastate}{Louisiana State University, Baton Rouge, LA 70803, USA}
\newcommand{\LpBordeaux}{Laboratoire de Physique des Deux Infinis Bordeaux - IN2P3, F-33175 Gradignan, Bordeaux, France, }
\newcommand{\Lucknow}{University of Lucknow, Uttar Pradesh 226007, India}
\newcommand{\Madrid}{Madrid Autonoma University and IFT UAM/CSIC, 28049 Madrid, Spain}
\newcommand{\Mainz}{Johannes Gutenberg-Universit{\"a}t Mainz, 55122 Mainz, Germany}
\newcommand{\Manchester}{University of Manchester, Manchester M13 9PL, United Kingdom}
\newcommand{\Massinsttech}{Massachusetts Institute of Technology, Cambridge, MA 02139, USA}
\newcommand{\Medellin}{University of Medell{\'\i}n, Medell{\'\i}n, 050026 Colombia }
\newcommand{\Michigan}{University of Michigan, Ann Arbor, MI 48109, USA}
\newcommand{\Michiganstate}{Michigan State University, East Lansing, MI 48824, USA}
\newcommand{\MilanoBicocca}{Universit{\`a} di Milano Bicocca , 20126 Milano, Italy}
\newcommand{\MilanoUniv}{Universit{\`a} degli Studi di Milano, I-20133 Milano, Italy}
\newcommand{\Minnduluth}{University of Minnesota Duluth, Duluth, MN 55812, USA}
\newcommand{\Minntwin}{University of Minnesota Twin Cities, Minneapolis, MN 55455, USA}
\newcommand{\Mississippi}{University of Mississippi, University, MS 38677 USA}
\newcommand{\napoli}{Universit{\`a} degli Studi di Napoli Federico II , 80138 Napoli NA, Italy}
\newcommand{\Nikhef}{Nikhef National Institute of Subatomic Physics, 1098 XG Amsterdam, Netherlands}
\newcommand{\Niser}{National Institute of Science Education and Research (NISER), Odisha 752050, India}
\newcommand{\Northdakota}{University of North Dakota, Grand Forks, ND 58202-8357, USA}
\newcommand{\Northernillinois}{Northern Illinois University, DeKalb, IL 60115, USA}
\newcommand{\Northwestern}{Northwestern University, Evanston, Il 60208, USA}
\newcommand{\NotreDame}{University of Notre Dame, Notre Dame, IN 46556, USA}
\newcommand{\NoviSad}{University of Novi Sad, 21102 Novi Sad, Serbia}
\newcommand{\Occidental}{Occidental College, Los Angeles, CA  90041}
\newcommand{\Ohiostate}{Ohio State University, Columbus, OH 43210, USA}
\newcommand{\OregonState}{Oregon State University, Corvallis, OR 97331, USA}
\newcommand{\Oxford}{University of Oxford, Oxford, OX1 3RH, United Kingdom}
\newcommand{\PacificNorthwest}{Pacific Northwest National Laboratory, Richland, WA 99352, USA}
\newcommand{\Padova}{Universt{\`a} degli Studi di Padova, I-35131 Padova, Italy}
\newcommand{\Panjab}{Panjab University, Chandigarh, 160014, India}
\newcommand{\Parissaclay}{Universit{\'e} Paris-Saclay, CNRS/IN2P3, IJCLab, 91405 Orsay, France}
\newcommand{\Parisuniversite}{Universit{\'e} Paris Cit{\'e}, CNRS, Astroparticule et Cosmologie, Paris, France}
\newcommand{\Parma}{University of Parma,  43121 Parma PR, Italy}
\newcommand{\Pavia}{Universit{\`a} degli Studi di Pavia, 27100 Pavia PV, Italy}
\newcommand{\Penn}{University of Pennsylvania, Philadelphia, PA 19104, USA}
\newcommand{\PennState}{Pennsylvania State University, University Park, PA 16802, USA}
\newcommand{\PhysicalResearchLaboratory}{Physical Research Laboratory, Ahmedabad 380 009, India}
\newcommand{\Pisa}{Universit{\`a} di Pisa, I-56127 Pisa, Italy}
\newcommand{\Pitt}{University of Pittsburgh, Pittsburgh, PA 15260, USA}
\newcommand{\Pontificia}{Pontificia Universidad Cat{\'o}lica del Per{\'u}, Lima, Per{\'u}}
\newcommand{\PuertoRico}{University of Puerto Rico, Mayaguez 00681, Puerto Rico, USA}
\newcommand{\Punjab}{Punjab Agricultural University, Ludhiana 141004, India}
\newcommand{\QMUL}{Queen Mary University of London, London E1 4NS, United Kingdom }
\newcommand{\Radboud}{Radboud University, NL-6525 AJ Nijmegen, Netherlands}
\newcommand{\Rice}{Rice University, Houston, TX 77005}
\newcommand{\Rochester}{University of Rochester, Rochester, NY 14627, USA}
\newcommand{\Royalholloway}{Royal Holloway College London, London, TW20 0EX, United Kingdom}
\newcommand{\Rutgers}{Rutgers University, Piscataway, NJ, 08854, USA}
\newcommand{\Rutherford}{STFC Rutherford Appleton Laboratory, Didcot OX11 0QX, United Kingdom}
\newcommand{\Salento}{Universit{\`a} del Salento, 73100 Lecce, Italy}
\newcommand{\santamarta}{Universidad del Magdalena, Santa Marta - Colombia}
\newcommand{\Sapienza}{Sapienza University of Rome, 00185 Roma RM, Italy}
\newcommand{\SergioArboleda}{Universidad Sergio Arboleda, 11022 Bogot{\'a}, Colombia}
\newcommand{\Sheffield}{University of Sheffield, Sheffield S3 7RH, United Kingdom}
\newcommand{\SLAC}{SLAC National Accelerator Laboratory, Menlo Park, CA 94025, USA}
\newcommand{\Southcarolina}{University of South Carolina, Columbia, SC 29208, USA}
\newcommand{\SouthDakotaSchool}{South Dakota School of Mines and Technology, Rapid City, SD 57701, USA}
\newcommand{\SouthDakotaState}{South Dakota State University, Brookings, SD 57007, USA}
\newcommand{\SouthernMethodist}{Southern Methodist University, Dallas, TX 75275, USA}
\newcommand{\StonyBrook}{Stony Brook University, SUNY, Stony Brook, NY 11794, USA}
\newcommand{\Sunyatsen}{Sun Yat-Sen University, Guangzhou, 510275, China}
\newcommand{\SURF}{Sanford Underground Research Facility, Lead, SD, 57754, USA}
\newcommand{\Sussex}{University of Sussex, Brighton, BN1 9RH, United Kingdom}
\newcommand{\Syracuse}{Syracuse University, Syracuse, NY 13244, USA}
\newcommand{\Tecnologica }{Universidade Tecnol{\'o}gica Federal do Paran{\'a}, Curitiba, Brazil}
\newcommand{\TelAviv}{Tel Aviv University, Tel Aviv-Yafo, Israel}
\newcommand{\TexasAMcollege}{Texas A{\&}M University, College Station, Texas 77840}
\newcommand{\TexasAMcorpuscristi}{Texas A{\&}M University - Corpus Christi, Corpus Christi, TX 78412, USA}
\newcommand{\TexasArlington}{University of Texas at Arlington, Arlington, TX 76019, USA}
\newcommand{\Texasaustin}{University of Texas at Austin, Austin, TX 78712, USA}
\newcommand{\Toronto}{University of Toronto, Toronto, Ontario M5S 1A1, Canada}
\newcommand{\Tufts}{Tufts University, Medford, MA 02155, USA}
\newcommand{\Unifesp}{Universidade Federal de S{\~a}o Paulo, 09913-030, S{\~a}o Paulo, Brazil}
\newcommand{\UNIST}{Ulsan National Institute of Science and Technology, Ulsan 689-798, South Korea}
\newcommand{\UniversityCollegeLondon}{University College London, London, WC1E 6BT, United Kingdom}
\newcommand{\UNMSM}{Universidad Nacional Mayor de San Marcos, Lima, Peru}
\newcommand{\ValleyCity}{Valley City State University, Valley City, ND 58072, USA}
\newcommand{\Vigo}{University of Vigo, E- 36310 Vigo Spain}
\newcommand{\VirginiaTech}{Virginia Tech, Blacksburg, VA 24060, USA}
\newcommand{\Warsaw}{University of Warsaw, 02-093 Warsaw, Poland}
\newcommand{\Warwick}{University of Warwick, Coventry CV4 7AL, United Kingdom}
\newcommand{\Wellesley}{Wellesley College, Wellesley, MA 02481, USA}
\newcommand{\Wichita}{Wichita State University, Wichita, KS 67260, USA}
\newcommand{\WilliamMary}{William and Mary, Williamsburg, VA 23187, USA}
\newcommand{\Wisconsin}{University of Wisconsin Madison, Madison, WI 53706, USA}
\newcommand{\Yale}{Yale University, New Haven, CT 06520, USA}
\newcommand{\Yerevan}{Yerevan Institute for Theoretical Physics and Modeling, Yerevan 0036, Armenia}
\newcommand{\York}{York University, Toronto M3J 1P3, Canada}
\affiliation{\Abilene}
\affiliation{\Albanysuny}
\affiliation{\Amsterdam}
\affiliation{\Antalya}
\affiliation{\Antananarivo}
\affiliation{\Antioquia}
\affiliation{\AntonioNarino}
\affiliation{\Argonne}
\affiliation{\Arizona}
\affiliation{\Asuncion}
\affiliation{\Athens}
\affiliation{\Atlantico}
\affiliation{\Augustana}
\affiliation{\Bern}
\affiliation{\Beykent}
\affiliation{\Birmingham}
\affiliation{\BolognaUniversity}
\affiliation{\Boston}
\affiliation{\Bristol}
\affiliation{\Brookhaven}
\affiliation{\Bucharest}
\affiliation{\CalBerkeley}
\affiliation{\CalDavis}
\affiliation{\CalIrvine}
\affiliation{\CalLosangeles}
\affiliation{\CalRiverside}
\affiliation{\CalSantabarbara}
\affiliation{\Caltech}
\affiliation{\Cambridge}
\affiliation{\Campinas}
\affiliation{\CataniaUniversitadi}
\affiliation{\Catolica}
\affiliation{\CBPF}
\affiliation{\CEASaclay}
\affiliation{\CERN}
\affiliation{\Charles}
\affiliation{\Chicago}
\affiliation{\ChungAng}
\affiliation{\CIEMAT}
\affiliation{\Cincinnati}
\affiliation{\Cinvestav}
\affiliation{\Colima}
\affiliation{\ColoradoBoulder}
\affiliation{\ColoradoState}
\affiliation{\Columbia}
\affiliation{\conida}
\affiliation{\Cti}
\affiliation{\CUSB}
\affiliation{\CzechAcademyofSciences}
\affiliation{\CzechTechnical}
\affiliation{\DannecyleVieux}
\affiliation{\Daresbury}
\affiliation{\Dordt}
\affiliation{\Drexel}
\affiliation{\Duke}
\affiliation{\Durham}
\affiliation{\Edinburgh}
\affiliation{\EIA}
\affiliation{\erciyes}
\affiliation{\Eotvos}
\affiliation{\FCULport}
\affiliation{\FederaldeAlfenas}
\affiliation{\FederaldeGoias}
\affiliation{\FederaldoABC}
\affiliation{\FederaldoRio}
\affiliation{\Fermi}
\affiliation{\Ferrarauniv}
\affiliation{\Florida}
\affiliation{\Floridastate}
\affiliation{\Fluminense}
\affiliation{\Genova}
\affiliation{\Georgian}
\affiliation{\Granada}
\affiliation{\GranSasso}
\affiliation{\GranSassoLab}
\affiliation{\Grenoble}
\affiliation{\Guanajuato}
\affiliation{\Harish}
\affiliation{\Hawaii}
\affiliation{\hkust}
\affiliation{\Houston}
\affiliation{\Hyderabad}
\affiliation{\Idaho}
\affiliation{\IFIC}
\affiliation{\IGFAE}
\affiliation{\Iitk}
\affiliation{\Illinoisinstitute}
\affiliation{\Imperial}
\affiliation{\IndGuwahati}
\affiliation{\IndHyderabad}
\affiliation{\Indiana}
\affiliation{\INFNBologna}
\affiliation{\INFNCatania}
\affiliation{\INFNFerrara}
\affiliation{\INFNFrascati}
\affiliation{\INFNGenova}
\affiliation{\INFNLecce}
\affiliation{\INFNMilanBicocca}
\affiliation{\INFNMilano}
\affiliation{\INFNNapoli}
\affiliation{\INFNPadova}
\affiliation{\INFNPavia}
\affiliation{\INFNPisa}
\affiliation{\INFNRoma}
\affiliation{\INFNSud}
\affiliation{\Ingenieria}
\affiliation{\Insubria }
\affiliation{\Iowa}
\affiliation{\IowaState}
\affiliation{\IPLyon}
\affiliation{\IPM}
\affiliation{\ISTlisboa}
\affiliation{\Ita}
\affiliation{\Iwate}
\affiliation{\Jacksonstate}
\affiliation{\Jawaharlal}
\affiliation{\Jeonbuk}
\affiliation{\Jyvaskyla}
\affiliation{\Kansasstate}
\affiliation{\Kavli}
\affiliation{\KEK}
\affiliation{\KISTI}
\affiliation{\Kure}
\affiliation{\Kyiv}
\affiliation{\Lancaster}
\affiliation{\LawrenceBerkeley}
\affiliation{\LIP}
\affiliation{\Liverpool}
\affiliation{\LosAlmos}
\affiliation{\Louisanastate}
\affiliation{\LpBordeaux}
\affiliation{\Lucknow}
\affiliation{\Madrid}
\affiliation{\Mainz}
\affiliation{\Manchester}
\affiliation{\Massinsttech}
\affiliation{\Medellin}
\affiliation{\Michigan}
\affiliation{\Michiganstate}
\affiliation{\MilanoBicocca}
\affiliation{\MilanoUniv}
\affiliation{\Minnduluth}
\affiliation{\Minntwin}
\affiliation{\Mississippi}
\affiliation{\napoli}
\affiliation{\Nikhef}
\affiliation{\Niser}
\affiliation{\Northdakota}
\affiliation{\Northernillinois}
\affiliation{\Northwestern}
\affiliation{\NotreDame}
\affiliation{\NoviSad}
\affiliation{\Occidental}
\affiliation{\Ohiostate}
\affiliation{\OregonState}
\affiliation{\Oxford}
\affiliation{\PacificNorthwest}
\affiliation{\Padova}
\affiliation{\Panjab}
\affiliation{\Parissaclay}
\affiliation{\Parisuniversite}
\affiliation{\Parma}
\affiliation{\Pavia}
\affiliation{\Penn}
\affiliation{\PennState}
\affiliation{\PhysicalResearchLaboratory}
\affiliation{\Pisa}
\affiliation{\Pitt}
\affiliation{\Pontificia}
\affiliation{\PuertoRico}
\affiliation{\Punjab}
\affiliation{\QMUL}
\affiliation{\Radboud}
\affiliation{\Rice}
\affiliation{\Rochester}
\affiliation{\Royalholloway}
\affiliation{\Rutgers}
\affiliation{\Rutherford}
\affiliation{\Salento}
\affiliation{\santamarta}
\affiliation{\Sapienza}
\affiliation{\SergioArboleda}
\affiliation{\Sheffield}
\affiliation{\SLAC}
\affiliation{\Southcarolina}
\affiliation{\SouthDakotaSchool}
\affiliation{\SouthDakotaState}
\affiliation{\SouthernMethodist}
\affiliation{\StonyBrook}
\affiliation{\Sunyatsen}
\affiliation{\SURF}
\affiliation{\Sussex}
\affiliation{\Syracuse}
\affiliation{\Tecnologica }
\affiliation{\TelAviv}
\affiliation{\TexasAMcollege}
\affiliation{\TexasAMcorpuscristi}
\affiliation{\TexasArlington}
\affiliation{\Texasaustin}
\affiliation{\Toronto}
\affiliation{\Tufts}
\affiliation{\Unifesp}
\affiliation{\UNIST}
\affiliation{\UniversityCollegeLondon}
\affiliation{\UNMSM}
\affiliation{\ValleyCity}
\affiliation{\Vigo}
\affiliation{\VirginiaTech}
\affiliation{\Warsaw}
\affiliation{\Warwick}
\affiliation{\Wellesley}
\affiliation{\Wichita}
\affiliation{\WilliamMary}
\affiliation{\Wisconsin}
\affiliation{\Yale}
\affiliation{\Yerevan}
\affiliation{\York}
\author{A.~Abed Abud} \affiliation{\CERN}
\author{B.~Abi} \affiliation{\Oxford}
\author{R.~Acciarri} \affiliation{\Fermi}
\author{M.~A.~Acero} \affiliation{\Atlantico}
\author{M.~R.~Adames} \affiliation{\Tecnologica }
\author{G.~Adamov} \affiliation{\Georgian}
\author{M.~Adamowski} \affiliation{\Fermi}
\author{D.~Adams} \affiliation{\Brookhaven}
\author{M.~Adinolfi} \affiliation{\Bristol}
\author{C.~Adriano} \affiliation{\Campinas}
% ...
\collaboration{The DUNE Collaboration}
\noaffiliation


\begin{abstract}
    The determination of the direction of a stellar core collapse via its neutrino emission is crucial for the identification of the progenitor for a multimessenger follow-up. A highly effective method of reconstructing supernova directions within the Deep Underground Neutrino Experiment (DUNE) is introduced. The supernova neutrino pointing resolution is studied by simulating and reconstructing electron-neutrino charged-current absorption on ${}^{40}\text{Ar}$ and elastic scattering of neutrinos on electrons. Procedures to reconstruct individual interactions, including a newly developed technique called ``brems flipping'', as well as the burst direction from an ensemble of interactions are described. 
    Performance of the burst direction reconstruction is evaluated for supernovae happening at a distance of 10\,kpc for a specific supernova burst flux model.
    The pointing resolution is found to be 3.4\,degrees at 68\% coverage for a perfect interaction-channel classification and a fiducial mass of 40\,kton, and 6.6\,degrees for a 10\,kton fiducial mass respectively. Assuming a 4\% rate of charged-current interactions being misidentified as elastic scattering, DUNE's burst pointing resolution is found to be 4.3\,degrees (8.7\,degrees) at 68\% coverage.
\end{abstract}

\maketitle

\tableofcontents

\section{Introduction}
 
Detecting neutrinos from core-collapse supernovae opens up the possibility to study the astrophysics of stellar collapse and the properties of neutrinos and their interactions.  The first and only supernova neutrino detection on Earth so far was achieved with SN1987A, when a few tens of electron-antineutrino events\footnote{We note here the use of ``event'' to refer to an individual recorded neutrino interaction, as per standard particle physics usage.  In this paper, ``burst" will refer to the ensemble of events from a single core collapse.} were registered~\cite{Bionta:1987qt,Hirata:1987hu,Alekseev:1987ej}. While it was already possible to derive valuable insights from these data (e.g., Ref.~\cite{burrows1987neutrinos,   schramm1990new}), a high-statistics, high-resolution detection of various neutrino flavors from a future supernova will bring invaluable advancement in particle physics and astrophysics.
Information can be derived from the detected neutrino event rate, timing, energy spectrum, flavor composition, and angular distribution.

An important characteristic feature of a supernova neutrino burst is that it emerges promptly after the core bounce, preceding the related electromagnetic phenomena. As neutrinos interact only via the weak interaction, they can escape more easily in comparison to photons, allowing the neutrino signal to be observed well before the associated electromagnetic radiation (as can gravitational radiation.) The Supernova Early Warning System (SNEWS)~\cite{al2021snews} is designed such that information retrieved from the supernova neutrino signal can be made available worldwide quickly, thereby facilitating the prompt detection of subsequent multi-messenger supernova-related phenomena~\cite{Adams:2013ana,Nakamura:2016kkl}.  

Clearly, it is highly desirable for information about the position of the supernova in the sky to be ascertained as fast as possible. Not only is the direction useful for the localization of the supernova to enable prompt measurements, but for the case when the core collapse fails to produce a bright explosion in electromagnetic radiation (e.g., black hole formation), a neutrino burst direction may help to locate a missing progenitor using archival data.  

Such pointing information is available from the neutrinos themselves. One possible strategy for determining the direction to the supernova is via ``triangulation" from the relative timing of neutrinos observed at different locations around the globe~\cite{Beacom:1998fj,Muhlbeier:2013gwa,Brdar:2018zds,Linzer:2019swe}. However, the most promising way to achieve precision pointing is to exploit anisotropic neutrino interactions~\cite{Beacom:1998fj} for which information about the incoming neutrino direction is preserved in the final-state particle angular distribution.  Water Cherenkov detectors such as Super-Kamiokande and Hyper-Kamiokande can make use of directional Cherenkov radiation to determine the supernova direction~\cite{abe2016real}.
 
We note also that for a known source direction, directional reconstruction of final-state particles can improve the neutrino energy determination, as well as statistical classification of interaction channels with known anisotropy.  Both of these can improve the extraction of physics and astrophysics from the burst.

The Deep Underground Neutrino Experiment (DUNE) is capable of a supernova burst detection among other physics goals~\cite{abi2021supernova,abi2020deep}.  An overview of DUNE's supernova detection capability is given in Ref.~\cite{abi2021supernova}.
In the several tens of kilotonnes (kton) of liquid argon (LAr) volume of DUNE, charged-current interactions of electron neutrinos ($\nu_e$CC) on ${}^{40}$Ar and neutrinos of all flavors scattering elastically on electrons (eES) will result in charge signals in DUNE's time projection chamber (TPC) and scintillation light that is read out via photosensors. 
In this work, both $\nu_e$CC and eES interactions are discussed.\footnote{We focus here on directional information derived from the expected prompt burst of tens-of-MeV neutrinos, although we note that higher energy (GeV scale or more) neutrinos may follow a supernova~\cite{Tomas:2003xn,Murase:2017pfe}. These may provide precision pointing due to both higher intrinsic lepton-neutrino collimation and better detector performance at high energy; however, they likely arrive on a long time scale (hours to years) after the supernova.} Primarily eES interactions carry directional information, but the near-isotropic $\nu_e$CC interactions have a higher cross section. A supernova burst alert from DUNE, including pointing information, will be a valuable input to SNEWS. 

The overall ability for DUNE to point to a supernova using the events recorded from a core collapse depends on several factors.  Two primary factors affect the direction resolution of individually recorded neutrino events.  First, there are intrinsic angular spreads between the recoil directions of the final-state products and the neutrino direction.  For eES interactions, which are the most important for the pointing ability, this energy-dependent spread is very well understood from weak interaction physics; for $\nu_e$CC it is less well known.  Second, the detector angular resolution smears the reconstructed direction with respect to the final-state electron direction. Detector-resolution smearing can in principle be improved with better reconstruction algorithms, although there will be an intrinsic physical limit for a given detector configuration. For this study, we use standard DUNE reconstruction techniques (detailed in Sec.~\ref{sec4}), with some minor improvements,  as well as a novel additional technique we call ``brems flipping'', which provides event-by-event a modest improvement in head-tail directional disambiguation by looking at the location of the secondary particles' tracks relative to the primary track.

While individual-event resolution on neutrino direction is relatively modest due to both physical angular spread and instrumental uncertainty, for the statistical ensemble of events in a burst the directional determination improves approximately by the inverse square root of the number of events detected.  We evaluate here the overall supernova burst pointing resolution, which will enable a meaningful contribution from DUNE to a multi-messenger detection of a core-collapse supernova.  The scope of this study, which makes use of standard offline reconstruction software, addresses only intrinsic pointing ability and does not consider latency for dissemination of pointing information.

Section~\ref{sec2} introduces the supernova neutrino-emission model that we use. In Sec.~\ref{sec3}, the DUNE detector and the expected signatures of the neutrino interaction channels in the LAr volume are described. Sec.~\ref{sec4} provides a description of the simulation of supernova neutrino events in DUNE and the subsequently applied reconstruction algorithms including the relevant software tools. This is followed by Sec.~\ref{sec5} on the maximum likelihood method for the burst direction determination. In Sec.~\ref{sec6} the overall performance of supernova burst direction reconstruction for DUNE is evaluated. Finally, Sec.~\ref{sec7} summarizes the results of the study and provides a road map for potential future studies in this area.

\section{Supernova neutrino emission}
\label{sec2}
 
When a massive star has depleted all of its nuclear fuel, the outgoing radiation pressure ceases to counteract the inward gravitational pull of the star, collapsing it into a compact object such as a neutron star or a black hole. During this process, ~99\% of the gravitational binding energy of the remnant is emitted in the form of neutrinos with energies of a few tens of MeV over a few tens of seconds. Supernova neutrinos are released in several stages during a core collapse. At the beginning of the collapse (over tens of milliseconds), neutrinos are primarily produced via electron capture ($p + e^- \rightarrow n + \nu_e$) as the star undergoes neutronization. During the subsequent accretion phase (which lasts for tens to hundreds of milliseconds), more neutrinos of all flavors are created, counteracting the shock heating of the in-falling matter. Subsequently, $\nu \bar{\nu}$ pairs are emitted over the next tens of seconds, such that these neutrino pairs shed most of the gravitational binding energy, thereby cooling the remnant \cite{scholberg2012supernova}. 

The study of supernova neutrinos is particularly useful in providing insights on varied topics in astrophysics and neutrino physics. As neutrinos are intimately involved in the collapse and subsequent explosion processes, measurements of the supernova neutrino signal allow for the examination of the complex interactions that occur during the core collapse. Additionally, supernova neutrino signals have the important feature that the initial luminosity is roughly equally divided among flavors.  The subsequent flavor transitions provide information on the neutrino mass ordering, as well as insights into exotic flavor transition physics \cite{duan2010collective,mirizzi2016supernova,gil2003oscillation,Scholberg:2017czd}.

For this study, the GKVM \cite{gava2009dynamical} core-collapse supernova neutrino emission model is used to describe the neutrino energy spectrum.  There are known to be fairly large (a factor of several) uncertainties on the supernova neutrino event rate prediction.  These uncertainties result both from intrinsic progenitor variance and from theoretical uncertainties.  These variances will have an impact on the total number of neutrinos detected as well as on details of the neutrino flavor composition and spectra.  However, the pointing capabilities of the DUNE detector are primarily sensitive to event statistics rather than to details of the model flux.  We therefore do not survey different supernova models here. The selected model produces an intermediate number of eES events among a range of models. Expected event rates without energy smearing are calculated with \snowglobes{} \cite{snowglobes}, which computes the recoil energy distribution as described in the next section. It is assumed that the supernova explosion occurs at a distance of 10\,kpc from Earth, within the Milky Way for our selected model. The total expected number of events and the events per interaction channel are presented in Tab.~\ref{tab:event_rates}. We do not study the effects of flavor transitions\footnote{Flavor transitions may affect the total number of neutrino events, but with uncertainty not exceeding the overall flux model uncertainties; spectral modulations will be a subdominant effect on pointing capabilities.}.

\section{Supernova neutrino detection at DUNE}
\label{sec3}

DUNE is unique in the sense that it will be able to register large numbers of electron neutrinos, while all other experiments of similar size, e.g., Hyper-K~\cite{abe2018hyper} and JUNO~\cite{an2016neutrino}, primarily detect electron antineutrinos via inverse beta decay on free protons. Immediately following the core collapse, electron-neutrino emission through neutronization dominates and insights on the neutrino mass ordering can be gained from observed differences in the neutrino flavor composition and spectra due to oscillation dynamics within the supernova~\cite{Scholberg:2017czd}.
 
We consider neutrino detection at DUNE's far detector liquid argon time projection chamber (LArTPC) \cite{abi2020volume}. The full planned design consists of 40\,kton of fiducial mass\footnote{We note this is the fiducial mass for long-baseline physics; active mass for supernova event detection could potentially be larger.} of liquid argon placed into four modules. Both single-module (10\,kton) and all-modules (40\,kton) performance is considered in this study. We consider the horizontal-drift module design for this study. Each of the active volumes is bound by the cathode plane assembly (CPA) and the anode plane assembly (APA) and surrounded by the field cage. A uniform electric field is created between the CPA and APA, drifting ionization charges toward the APA. Three planes of sensing wires, each with a different orientation, are located at the APA. Charge depositions on these readout planes are used to reconstruct the location of the particle energy deposition in two dimensions, and the drift timing information allows for the reconstruction of the third spatial coordinate. In addition to the LArTPCs, the Photon Detection System (PDS) is used to tag neutrino events. In this study, information on the timing of the interaction is retrieved from the detection of photon flashes to be able to estimate the charge loss during drift. More details about the photon detectors can be found in \cite{Falcone:20220h}.
 
The two neutrino-interaction channels considered in this study are neutrino-electron elastic scattering interactions and $\nu_e + \arforty$ CC absorption interactions. 
Neutral-current and anti-neutrino CC interactions are also expected~\cite{abi2021supernova}. However, as these channels are known to be subdominant and are at present not well understood, they are not included in this study. In the following, the nature of the dominant interaction channels is described and the simulated energy spectra and angular distributions in the detector are discussed. For eES events, the event rates are determined from \snowglobes, using the GKVM supernova model. It also provides the recoil energy distribution as input to a \larsoft{}~\cite{snider2017larsoft} neutrino-electron scattering event generator. For $\nu_e$CC events, MARLEY~\cite{gardiner2021simulating} and its \larsoft{} interface are applied to generate the events with the correct energy distribution. Once again the GKVM supernova model is provided as input to MARLEY in this case. All simulations are done in a subset of the full DUNE far detector geometry simulating 1.6\,kton of fiducial volume to reduce memory, disk, and computing time requirements.
The simulated workspace features six planes of APAs (two APAs tall) along the neutrino beam direction. A beam of high-energy neutrinos will be used for studies of neutrino oscillations with DUNE. For the full detector module, there are 150 APAs proposed in total, stacked two-tall and 25 along the beam direction. Because the spatial extent of the neutrino events under consideration is much smaller than the simulated workspace size, we expect the scaling down of the simulated geometry to have a negligible effect on the conclusions of this study.

\begin{table}[h]
    \centering
    \caption{The eES and $\nu_e$CC event rates (total for the interaction channel in bold) for DUNE within a fiducial volume of 40\,kton, and for a core collapse at a distance of 10\,kpc, using the GKVM model. No triggering efficiency or detector resolution effects are applied here, in contrast to \cite{abi2021supernova}. Neutrino flavor transition effects beyond those included in the GKVM model are neglected. The neutrino event rate scales linearly with detector mass and with inverse-square dependence on supernova distance.}
    \begin{ruledtabular}
    \begin{tabular}{ll}
    Channel & Expected Event Count\\
    \hline
    $\nu + e^-\rightarrow \nu + e^-   $ (all flavors)  &  \bf{325.8} \\
    $\nu_e + e^-\rightarrow \nu_e + e^-   $ & 155.5\\
    $\bar{\nu}_e + e^-\rightarrow \bar{\nu}_e + e^-   $ & 67.3 \\
    $\nu_{\mu, \tau} + e^-\rightarrow \nu_{\mu, \tau} + e^-   $ & 55.2\\
    $\bar{\nu}_{\mu, \tau} + e^-\rightarrow \bar{\nu}_{\mu, \tau} + e^-   $ & 47.8 \\
    $\nu_e + {}^{40}\text{Ar}\rightarrow e^- + {}^{40}\text{K}^*   $  &  \bf{3300.0}\\
    \end{tabular}
    \end{ruledtabular} 
    \label{tab:event_rates}
\end{table}

\subsection{Neutrino-Electron Elastic Scattering}
\begin{figure*}[htb]
    \centering
    \includegraphics[width=0.95\textwidth]{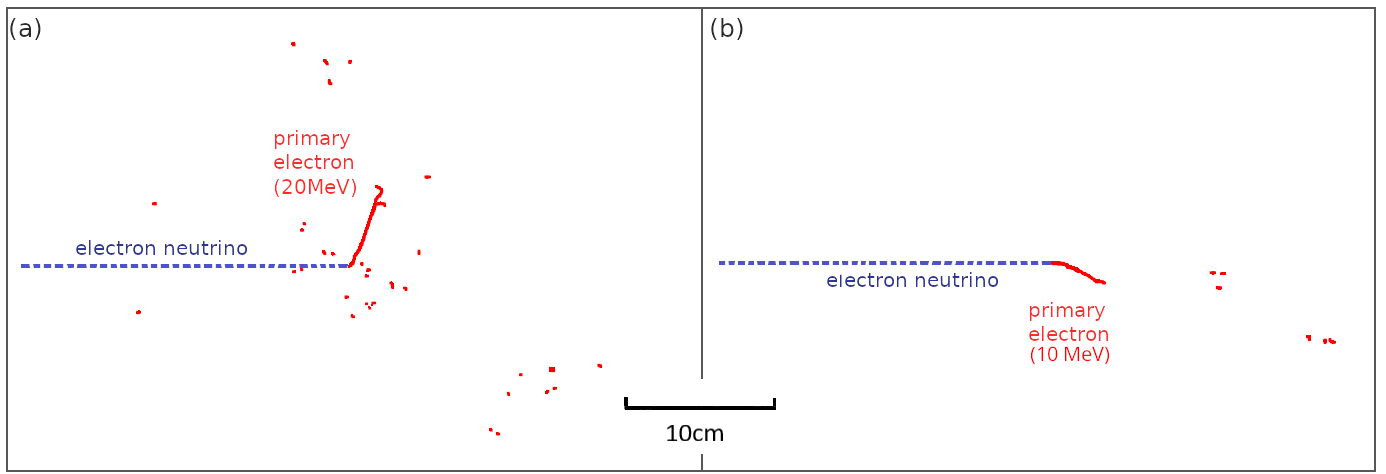}
    \caption{Geant4 illustration of the energy deposition for examples of two event types: (a) $\nu_e$CC: A 25\,MeV electron neutrino is absorbed by an argon nucleus resulting in the excitation of the nucleus and the emission of an electron. (b) eES: An incoming electron neutrino of 12\,MeV scatters elastically in the LAr. Bremsstrahlung gammas are preferentially emitted in the forward direction of the primary electron. The $\nu_e$CC primary electron tracks are on average longer than eES tracks due to the lower energies of the eES recoils; in contrast, $\nu_e$CC electrons tend to retain most of the energy of the incoming neutrino.
    Gamma tracks are not shown in the display, but red blips (representing electrons) from gamma interactions with the argon (primarily Compton scatters) can be seen.}
    \label{fig:geant4_event_display}
\end{figure*}

The most relevant interaction for directional information is when a neutrino elastically scatters off of an electron in the LAr. This interaction applies to all flavors but the largest cross section is for $\nu_e$. A visualization of a Geant4 event is shown in Fig.~\ref{fig:geant4_event_display} (b).

The direction of the scattered electron is highly correlated to the direction of the neutrino. In particular, the scattering angle $\theta_e$ is described by \cite{vogel1989neutrino}:
\begin{equation}
    \cos{\theta_e} = \frac{E_\nu + m_e}{E_\nu} \sqrt{\frac{T}{T+2m_e}},
    \label{eq:ES_kin}
\end{equation}

where $T$ is the electron kinetic energy, $E_\nu$ is the neutrino energy, and $m_e$ is the electron mass.
The distribution of $T$ is given by the differential cross section:
\begin{eqnarray}
    \dv{\sigma^{(\nu e)}}{T} &=& 
    \frac{{G_F}^2 m_e}{2\pi}\Biggl[(g_V + g_A)^2 \nonumber\\
    &&+ (g_V - g_A)^2\left(1-\frac{T}{E_\nu}\right)^2 \nonumber\\&&+ ({g_A}^2 -  {g_V}^2)\frac{m_e T}{{E_\nu}^2} \Biggr], 
    \label{eq:ES_xscn}
\end{eqnarray}

where $G_F$ is the Fermi coupling constant, and $g_A$ and $g_V$ are given according to neutrino flavor~\cite{vogel1989neutrino} in Tab.~\ref{tab:ES_params}.

\begin{table}[h]
\centering
\caption{Coupling strengths in the cross section of neutrino-electron scattering from Ref.~\cite{vogel1989neutrino}. The cross section of $\nu_e$ electron scattering is enhanced due to the possibility of both neutral- and charged-current interactions occurring; $\bar{\nu}_e$-electron elastic scattering is helicity-suppressed with respect to $\nu_e$ scattering.
}
\begin{ruledtabular}
\begin{tabular}{lll}
    Flavor & $g_A$ & $g_V$ \\
    \hline
    $\nu_e$ & $1/2$ & $2\sin^2{\theta_W} + 1/2$\\
    $\bar{\nu}_e$ & $-1/2$ & $2\sin^2{\theta_W} + 1/2$\\
    $\nu_{\mu, \tau}$ & $-1/2$ & $2\sin^2{\theta_W} - 1/2$\\
    $\bar{\nu}_{\mu, \tau}$ & $1/2$ & $2\sin^2{\theta_W} - 1/2$\\
\end{tabular}
\end{ruledtabular}
\label{tab:ES_params}
\end{table}

The event rate split into the different neutrino flavors can be found in Tab.~\ref{tab:event_rates}. The energy distributions of the generated neutrinos and scattered electrons are shown in Fig.~\ref{fig:ES_dist}(a, blue). The distribution of the scattering angle, computed via the energy distribution and Eq.~\ref{eq:ES_kin}, is depicted in Fig.~\ref{fig:ES_dist}(b, blue) showing the sensitivity to the direction of the incoming neutrino.

\subsection{Electron-Neutrino Charged-Current Absorption Interactions}
DUNE is particularly sensitive to the charged-current absorption of $\nu_e$ on $\arforty{}$ ($\nu_e$CC): 
\begin{equation}
    \nu_e + {}^{40}\text{Ar}\rightarrow e^- + {}^{40}\text{K}^*.
\end{equation}

\begin{figure*}[t]
    \centering
    \subfigure[]{\includegraphics[width=\columnwidth]{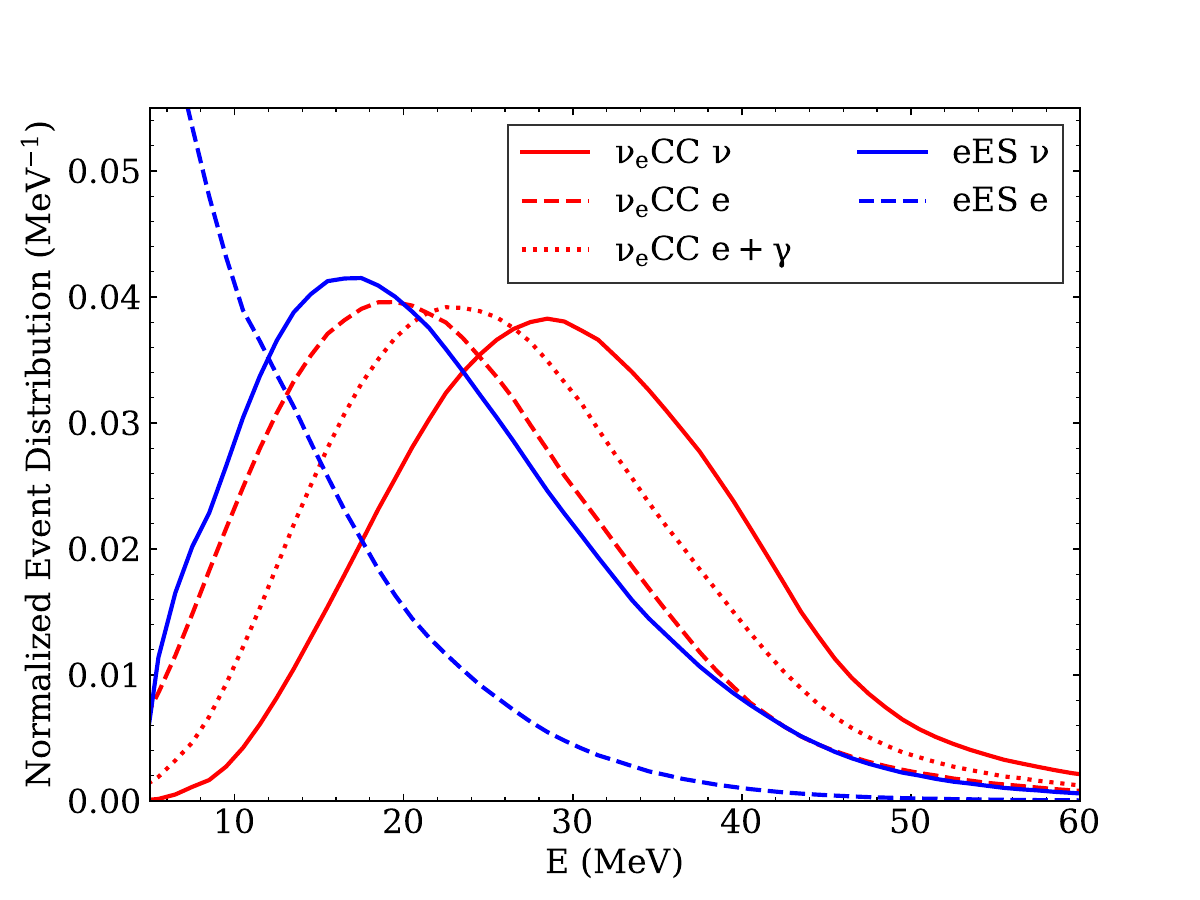}}
    \subfigure[]{\includegraphics[width=\columnwidth]{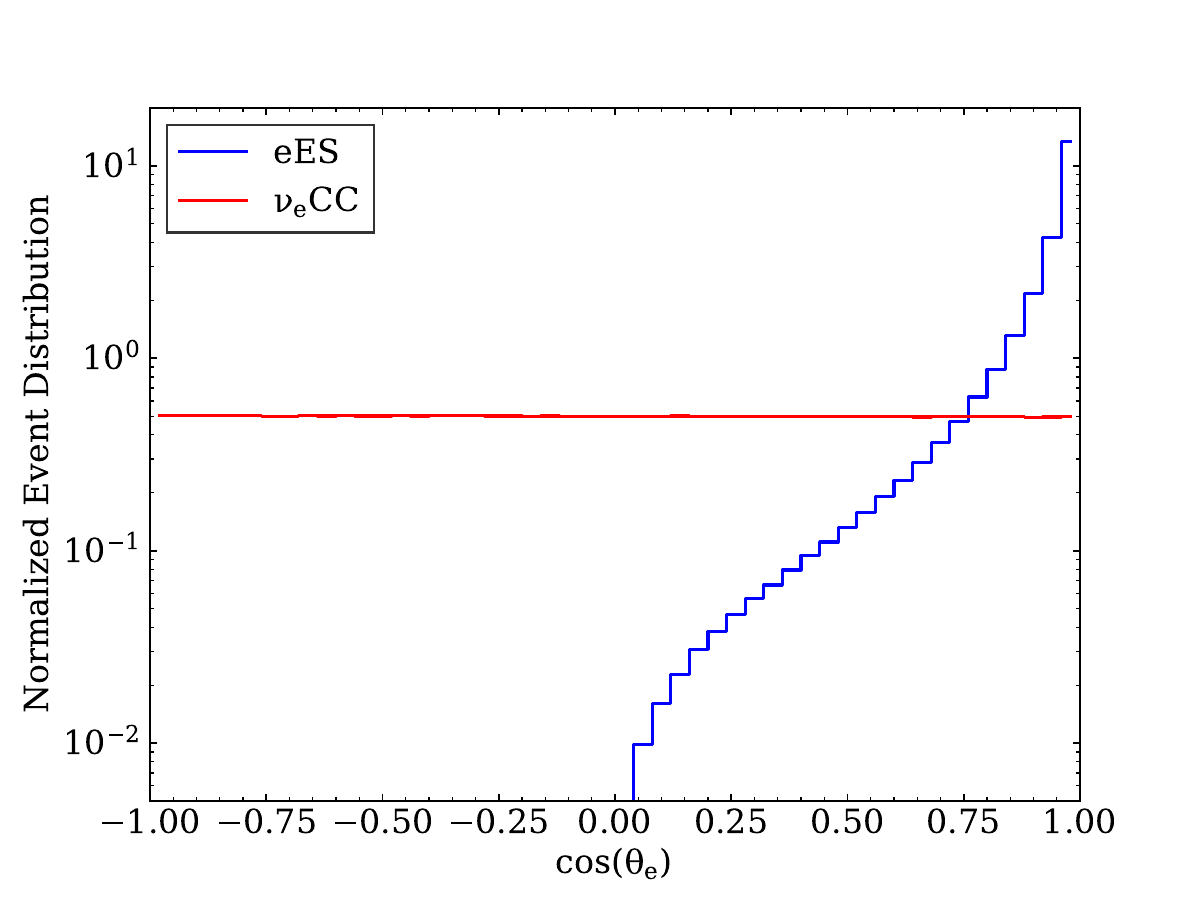}}
    \subfigure[]{\includegraphics[width=\columnwidth]{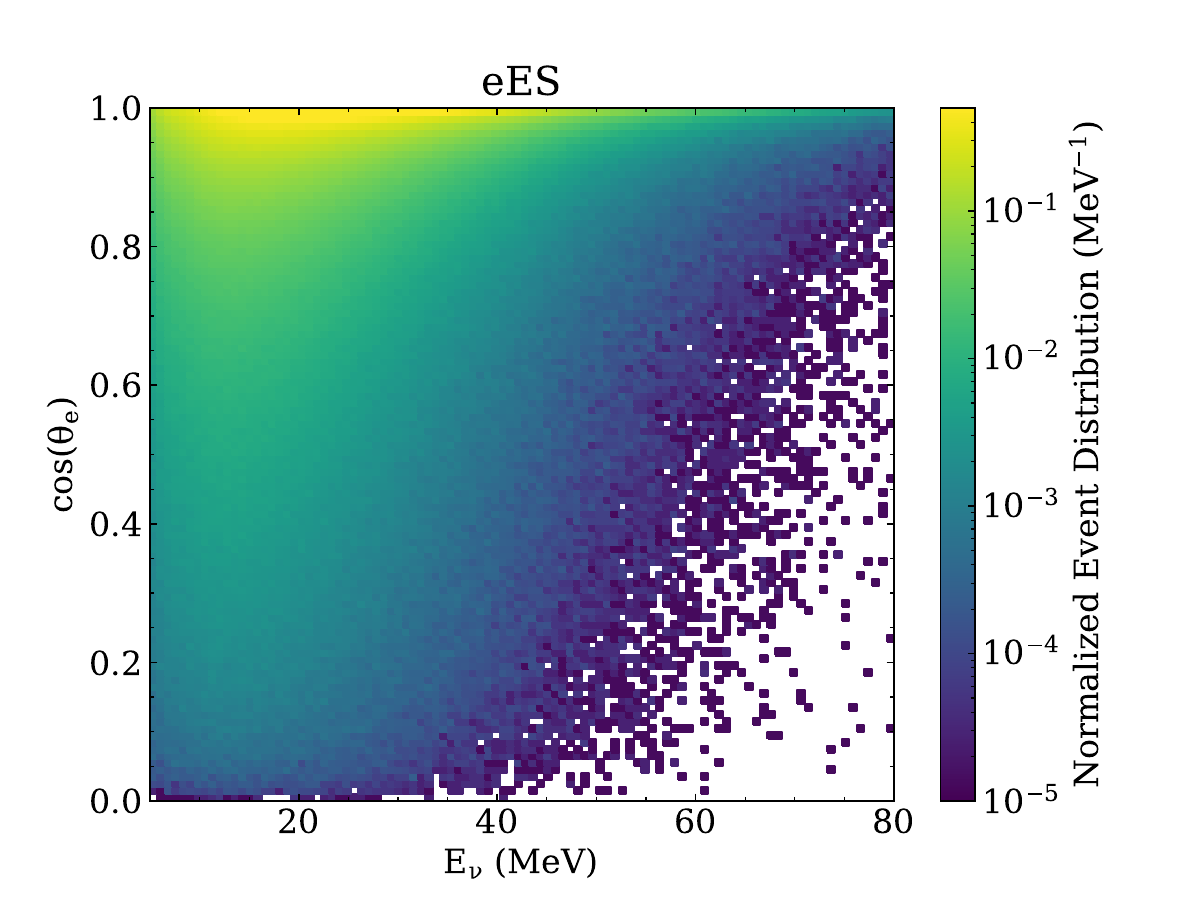}}
    \subfigure[]{\includegraphics[width=\columnwidth]{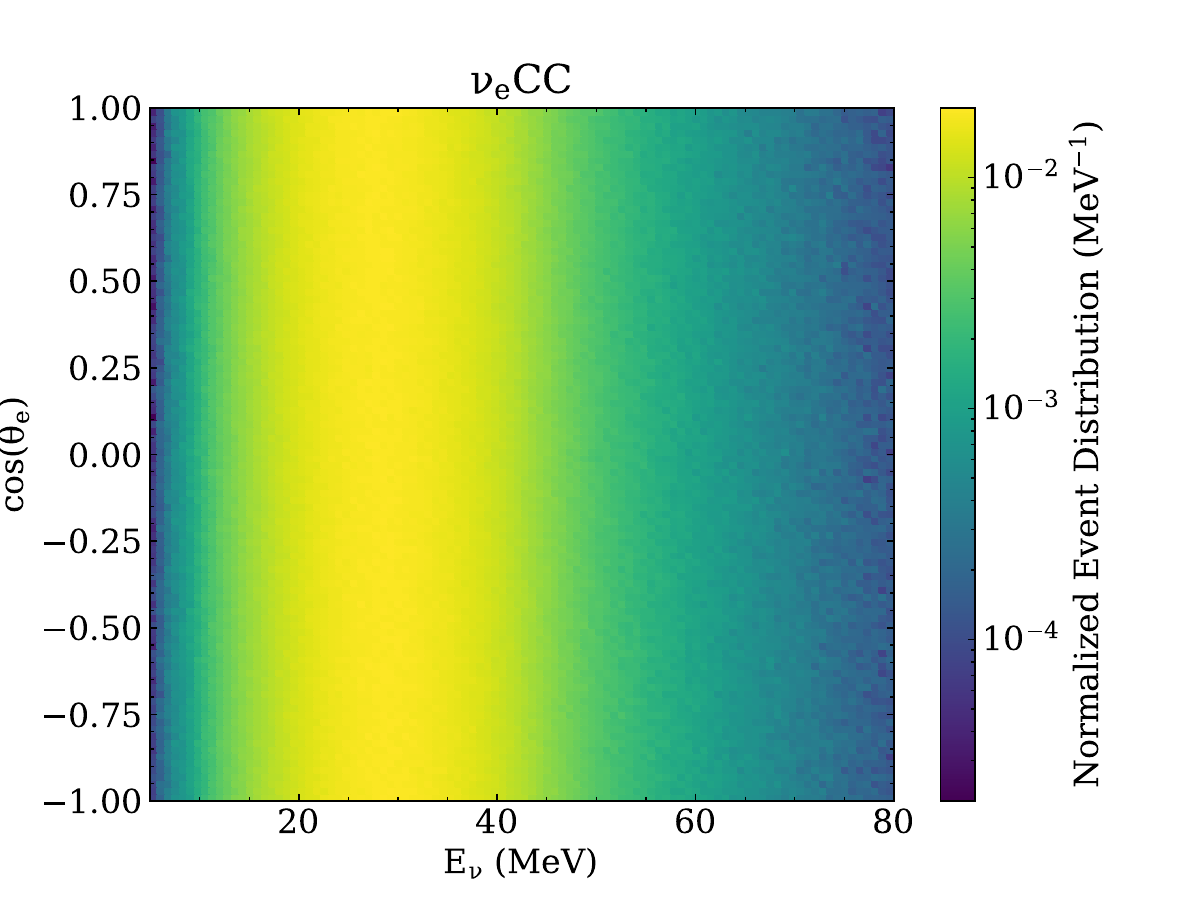}}
    \caption{Comparison of the cross-section-weighted energy and angular distributions of electrons for the $\nu_e + \prescript{40}{}{} \text{Ar}$ $\nu_e$CC events in red and the eES events in blue.  The angular plots are shown as a function of the cosine of the angle between the neutrino direction and the final-state electron direction.  For both types of interactions, the supernova energy spectrum is assumed for the incoming neutrinos. (a): distribution of the energy of incoming neutrinos (continuous line) and outgoing electrons (dashed line). (b): distribution of the cosine of the scattering angle. (c) and (d): two-dimensional distributions showing the relative number of events as a function of both the neutrino energy and the cosine of the scattering angle, for the two types of interactions.}
    \label{fig:ES_dist}
    \label{fig:CC_dist}
\end{figure*}

The primary observable of this interaction is the emitted $e^-$;  additional observables are the de-excitation products of the excited potassium nucleus in the final state. The output energy and angular distributions of this interaction are also shown in Fig.~\ref{fig:CC_dist} (a) and (b) in red and Fig.~\ref{fig:geant4_event_display} (a) depicts an example $\nu_e$CC event from Geant4.

\begin{figure}[htb]
    \centering
    \includegraphics[width=\columnwidth]{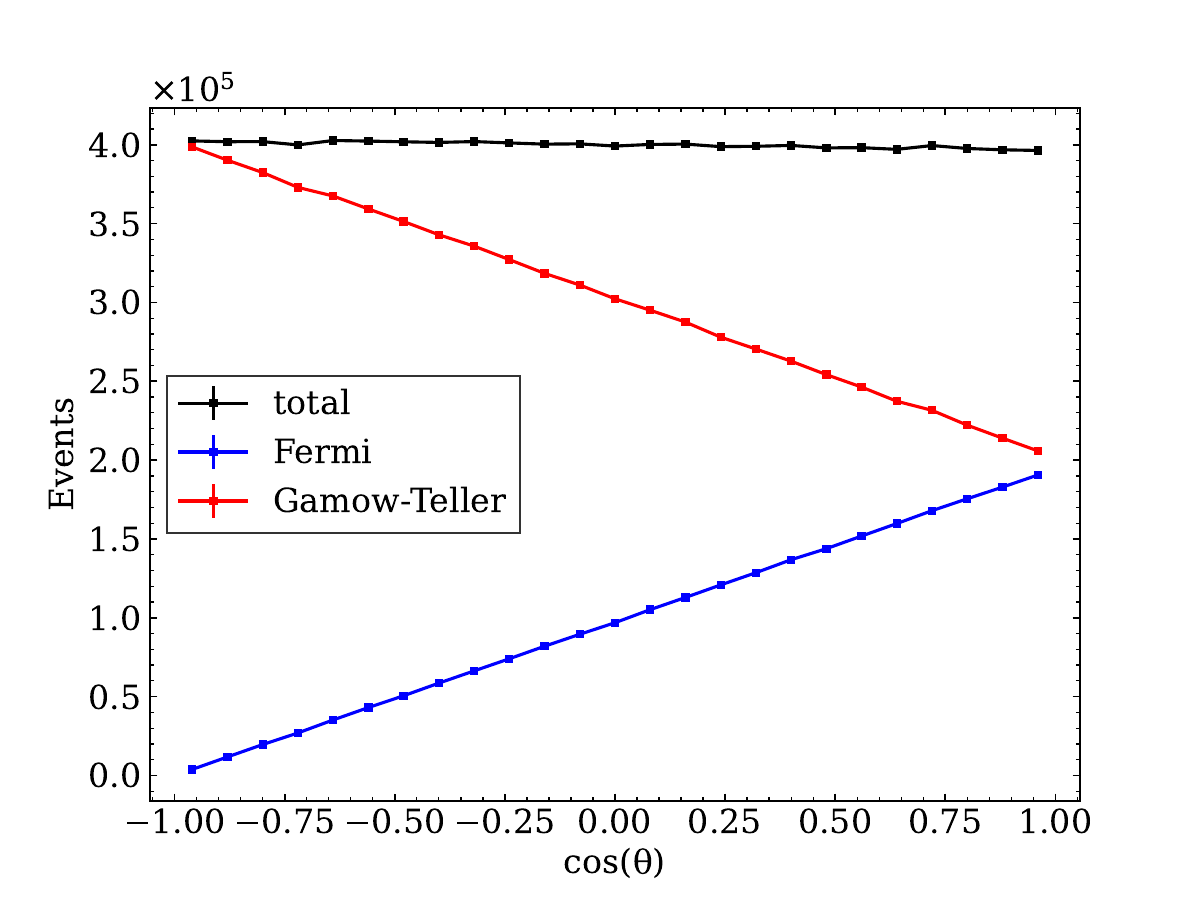}
    \caption{The contribution of Fermi and Gamow-Teller transitions to the angular distribution for $\nu_e$CC events, generated from MARLEY for the GVKM flux model. For this model's neutrino spectrum, the angular correlation cancels out to a good approximation.}
    \label{fig:CC_matrix_elements}
\end{figure}

Events from $\nu_e$CC interactions are considerably more abundant than the aforementioned eES events (about 3000 events at a core collapse distance of 10\,kpc for DUNE compared to 300 for eES in Tab.~\ref{tab:event_rates}.) Yet, the direction of the electron trajectory correlates very weakly with the neutrino direction, due to the two competing nuclear transition types, Fermi and Gamow-Teller. The electrons produced by CC-induced interactions governed by Fermi transitions have an angular distribution described by $1 + \frac{v}{c}\cos\theta$, while Gamow-Teller transitions correspond to $1 - \frac{1}{3}\frac{v}{c}\cos\theta$ \cite{gardiner2021nuclear} with respect to the neutrino direction. For our assumed supernova neutrino energy spectrum and cross-section model, the angular dependences of the two contributions happen to cancel each other out almost exactly, resulting in a nearly isotropic angular distribution for $\nu_e$CC events, as shown in the red distribution in Fig.~\ref{fig:CC_dist}(b). Figure \ref{fig:CC_matrix_elements} depicts the contribution of the two matrix elements for the GKVM supernova spectrum.  
The pointing resolution for a supernova burst therefore depends on a precise classifier between eES and $\nu_e$CC events as discussed in Sec.~\ref{sec5}. 

We note that the Fermi/Gamow-Teller cancellation is not perfect for other assumed flux spectra (e.g., Ref.~\cite{gardiner2021nuclear}), resulting in a weak anisotropy of the $\nu_e$CC-absorption-induced electrons for many cases.   Furthermore, forbidden transitions, not currently included in MARLEY, have a backward angular distribution~\cite{VanDessel:2019obk} which may have an effect on the overall $\nu_e$CC anisotropy.  Currently, there are large (and not fully understood) uncertainties on the relative components of $\nu_e$CC nuclear transition type and hence on the  $\nu_e$CC angular distribution  (as well as on the total cross section~\cite{DUNE:2023rtr}.)
We note that while the expected $\nu_e$CC anisotropy is relatively weak, it should be possible to extract directional information from the $\nu_e$CC events, especially if the competing nuclear transition types can be tagged and examined separately. In principle, the nuclear transition types can be distinguished statistically using their different nuclear de-excitation product patterns\footnote{This approach is under study for future application.}.   However, given the large uncertainties on the  $\nu_e$CC angular distribution\footnote{Note that these uncertainties may be addressed directly with independent laboratory measurements.} and that our assumed spectrum produces the most experimentally challenging situation--- a directional signal on top of a uniform $\nu_e$CC distribution--- we evaluate the pointing ability under this simple assumption and leave detailed study and use of $\nu_e$CC anisotropies for the future.

\section{Simulation and reconstruction of supernova events}
\label{sec4}

\subsection{Simulation of events in DUNE}
Full-detector Monte Carlo simulations are used to determine the statistical distribution of the particles resulting from the neutrino interactions and ultimately the supernova pointing resolution. 
 
The eES and $\nu_e$CC event generators in \larsoft{} produce particles uniformly in the workspace geometry with random neutrino directions sampled from an isotropic angular distribution. Standard radioactive and detector noise models were used during the simulations \cite{abi2020deep}. The considered radioactive background sources include contaminants of the liquid argon such as $^{39}$Ar and $^{85}$Kr, radioactive contaminants from the TPC, and neutrons from the cavern walls. We expect radiological and cosmogenic background to have a relatively small effect on the pointing quality. The dominant $^{39}$Ar background of 1\,Bq/kg results in only of the order of one background-induced blip expected in the spatial vicinity of a supernova event over one drift time~\cite{abi2020volume}. Therefore, we expect our evaluation to be robust against evolutions of the DUNE background model. Furthermore, the background will be known precisely; it can be fully characterized from the data taken near in time to the supernova burst, which will enable optimization of background mitigation for reconstruction algorithms. The simulated electronics signal process accounts for the charge deposition and drift physics, the field response of the sensing wires, electronics and digitization response of the front-end electronics, as well as the inherent electronics noise according to Ref.~\cite{abi2020volume}. The Projection Matching Algorithm~\cite{Antonello:2012hu} is used for 3D track reconstruction. 

\begin{figure}[htb]
    \centering
    \includegraphics[width=0.5\textwidth]{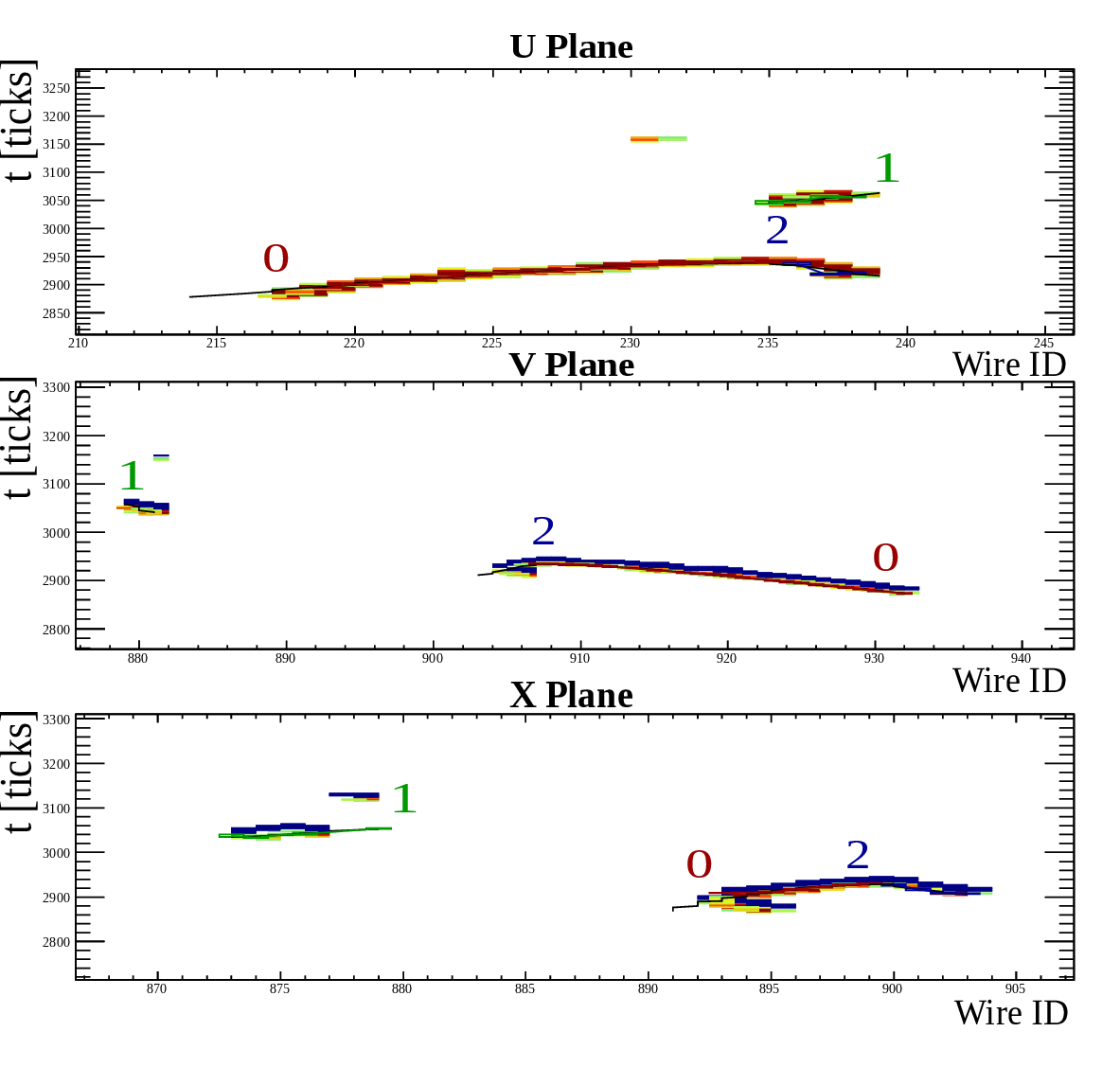}
    \caption{Example standard DUNE event display of an eES event. The colors show a heat map of the charge deposited on each wire segment per time tick. From top to bottom, the three plots show the view of the three wire planes, U, V, and X respectively, for one APA of the LArTPC. The trajectories with numerical labels are reconstructed tracks, where ``0'' corresponds to the primary electron track, while ``1" and ``2" refer to the bremsstrahlung-induced blips. }
    \label{fig:evd_example}
\end{figure}

An example eES event display is shown in Fig.~\ref{fig:evd_example}. Of note, the longest track marked by ``0" (several tens of cm) with high charge deposition corresponds to the primary electron and there are as well as several shorter tracks. These tracks result from lower energy final-state particles that were produced from the interaction (such as the deexcitation products of $\nu_e$CC events), or secondary particles created as the primary electron travels in the detector. Notably, some secondary particles have directions that correlate to the primary electron, making them useful in determining the starting direction of the primary electron track (discussed in Sec.~\ref{sec:reco}.) Additionally, charge may also be registered as a result of background radioactive decay events, such as the decay of $^{39}\text{Ar}$ and electronics noise. These signals are very short compared to the primary electron, are far apart from each other, and are of low amplitude, which are all characteristics that can be used for extracting signals from background.  

\subsection{Primary track identification and energy reconstruction}
\label{sec:reco}
The reconstruction algorithm consists of several steps.
First, the energy depositions associated with the supernova neutrinos are identified. The first step in reconstructing the event is to determine which reconstructed track corresponds to the primary electron (marked with ``0'' in Fig.~\ref{fig:evd_example}.) The DUNE version of the projection matching algorithm as described in \cite{Antonello:2012hu} is applied for the track reconstruction. We show in the following that the overall performance of the low-energy reconstruction is sufficient for the task at hand. Specific mitigations for the intrinsic head-tail ambiguity of the track will be described. In a noise- and background-free scenario, the track of the primary electron typically has the greatest length (see Fig.~\ref{fig:geant4_event_display},) as the electron deposits more charge compared to its secondary particles. However, the presence of radioactive decay particles can complicate this, as a highly energetic radioactive decay product may deposit more charge than the primary electron and create a longer track. These scenarios are rejected by examining the spatial distance between tracks. Particles associated with the neutrino interactions are clustered together, while the radioactive particles are spaced out over the total simulated LAr volume. Therefore, the primary track is selected by examining the relative position of charge depositions. Reconstructed 3D space points are sorted by their corresponding hit charges, and for the ten space points with the highest charge depositions, the distances between these space points are calculated. The two space points with the closest spatial distance between each other among those ten space points are picked and one of them is selected as the estimated reconstructed vertex position. While this location may not exactly be the true vertex, it allows the rough localization of the position inside the detector. The track with the most associated charge in the proximity of this reconstructed vertex (within a sphere of radius of 120\,cm) is finally selected to be the primary electron track.

Next, the energy of the primary electron is evaluated, as the pointing resolution of an event depends on the energy of its electron. Energy reconstruction is done by summing the total charge deposited near the identified interaction vertex (with a distance cut applied at 70\,cm, five times the 14\,cm radiation length of liquid Ar), and mapping charge to energy using a linear relationship. This distance cut is applied to reject the energy depositions of radioactive particles far away from the interaction vertex. Charge loss due to drift is also corrected by examining the time difference between the interaction time provided by the optical detector and the TPC signal collection time. An electron lifetime of 3000\,$\mu$s in LAr is assumed. Comparing the reconstructed primary electron energy after drift-time correction to the true energy of the simulated particles shows a linear relationship.

\subsection{Direction reconstruction and head-tail disambiguation}

Most relevant for the supernova pointing is the primary electron's direction. The goal is to attain an accurate ``reconstruction resolution'', which goes along with minimizing the sky area that must be surveyed to confidently encompass the supernova.   We expect a head-tail ambiguity in LArTPC track directions as the charge drift speed towards the anode is slow compared to the propagation of particles through the TPC material. Consequently, the track's head-tail direction cannot be determined through timing. In other words, it is not known through timing which side of the track corresponds to the origin of the track, causing the reconstructed direction to potentially be opposite of the true charged-particle track direction. This results in a bimodal distribution of angle between the true and reconstructed directon around the true and reversed directions of the primary track.  The reconstruction algorithm aims to enhance the peak pointing towards the true direction. 

\begin{figure}
    \centering
    \includegraphics[width=\linewidth]{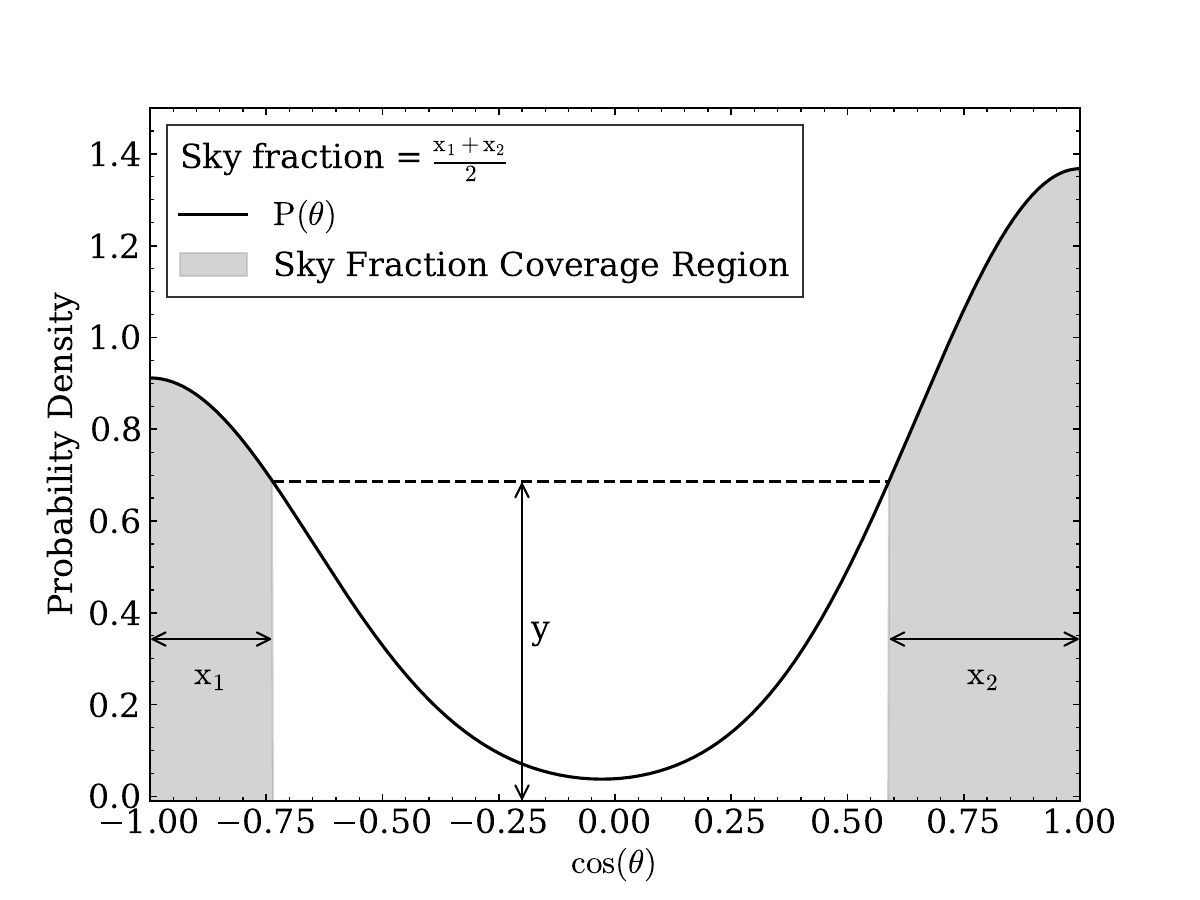}
    \caption{A graphical illustration of the sky fraction calculation for an example probability function in $\theta$. The sky fraction is the summed widths of the shaded regions, $x_1 + x_2$ divided by the domain width of $P(x)$, which is 2. The integration bounds $x_1$ and $x_2$ are set at the point where $P(x)$ crosses value $y$. The cut-off value $y$ is lowered until the integral of the shaded region equals 68\% of the total integral of $P(x)$.}
    \label{fig:skyfraction}
\end{figure}

To evaluate the performance of the reconstruction resolution, we consider detection probabilities from scanning the sky regions with the highest likelihood values first.   A histogram is filled with the distribution of cosine angles between the reconstructed direction and the truth direction. The histogram is integrated from both the forward and backward directions inwards for all probability densities exceeding a threshold; the threshold is lowered until the integral equals 68\% of the total integral of the distribution. The resolution metric ``sky fraction'' is defined to be the fraction of the sky that is covered by the 68\% integration.  Calculation of the sky fraction is graphically illustrated in Fig.~\ref{fig:skyfraction}. The sky fraction converts to an equivalent area of coverage in solid angle by multiplying by $4\pi$. For a small-angle approximation, the solid angle corresponds to $\pi$ times the opening angle squared, which can be considered the pointing resolution. Similarly, multiplying the sky fraction by two yields an effective confidence interval in $\cos{\theta}$. If the two sides of the distribution are close to fully symmetric, the sky fraction would be twice as large as the fully unambiguous case, corresponding to an equal probability of success for finding the supernova on each side of the sky.

The head-tail ambiguity can be resolved statistically by looking at the adjacent tracks created by secondary particles. The bremsstrahlung gamma rays from the primary electron subsequently release electrons in argon atoms via Compton scattering. These secondary electrons correlate with the forward direction of the primary electron. The directional correlation is closer when the primary electron is of higher energy. The particular procedure of disambiguating the direction of the primary electron is carried out in a process called ``brems flipping''.  
To apply the method, a starting-point and an end-point of the primary electron track are assumed arbitrarily. The vectors from the starting point as well as from the end point of the primary track to each secondary track are determined. Subsequently, the cosine value of the angle between each vector pointing to the secondary tracks and the vector along the primary track is evaluated. The average of these cosine values is calculated.  
In the next step, the assignment of the starting-point and the end-point are switched and the same calculations are carried out. Of the two vertices, the one with the larger average cosine value is finally selected as the actual starting point of the primary track. Most of the secondary particles are emitted towards the end of the primary track in a forward direction, leading to a preference for larger cosine values (corresponding to smaller angles) as can be understood from Fig.~\ref{fig:daughter_flipping_demo}. The brems-flipping method is also described in Algorithm \ref{alg:df} (see Appendix).

\begin{figure}[htb]
    \centering
    \includegraphics[width=0.53\textwidth]{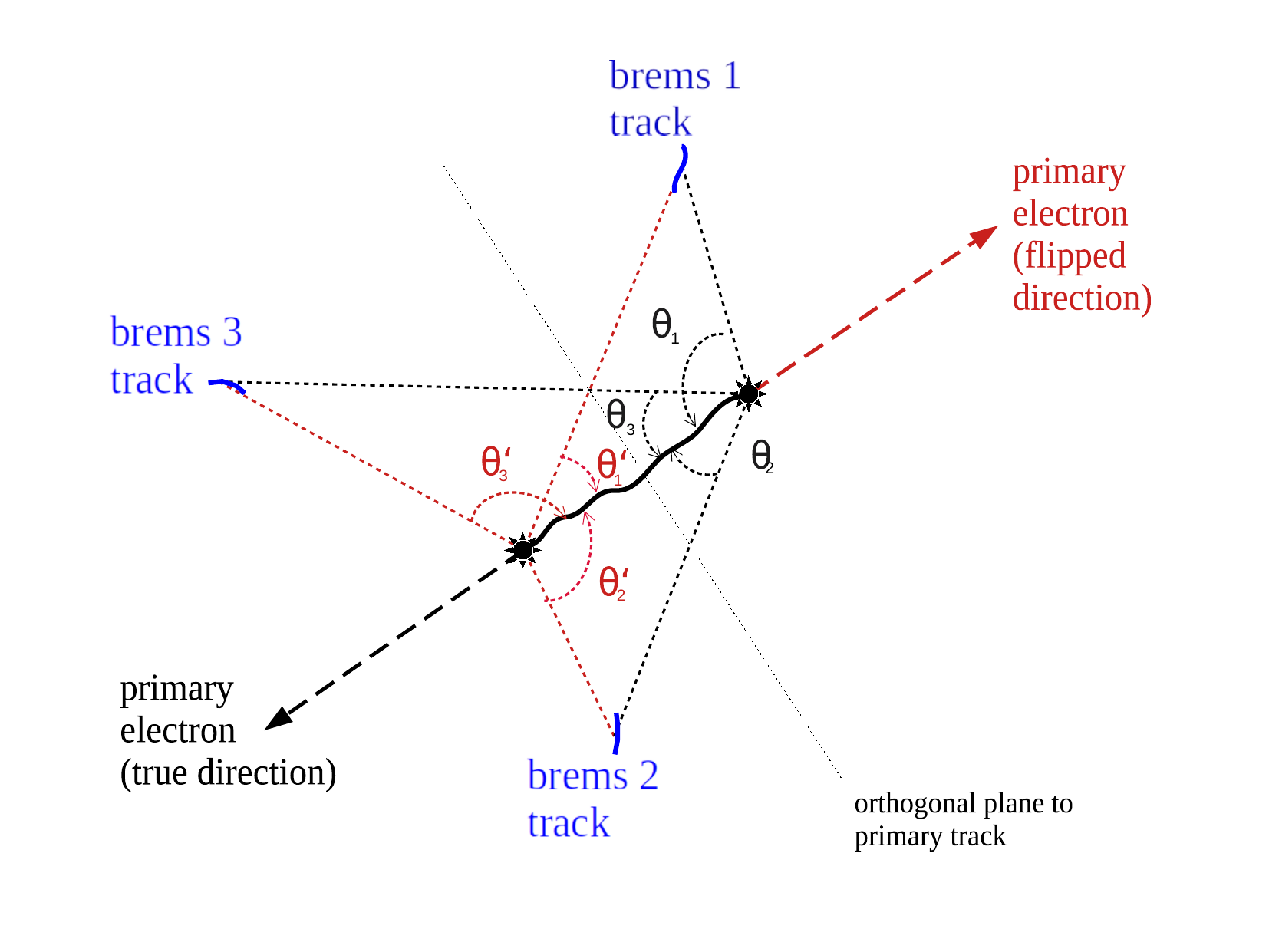}
    \caption{Illustration of brems flipping: The angles between the brems particles (blue) and the primary track marked in black correspond to the actual direction of the primary electron, while the ones marked in red belong to the incorrect opposite direction. The average of the cosine of these angles is larger in the case of the correct set of angles.}
    \label{fig:daughter_flipping_demo}
\end{figure}

\subsection{Performance of the reconstruction algorithm on single events}

\begin{figure}[h!]
    \centering
    \includegraphics[width=\columnwidth]{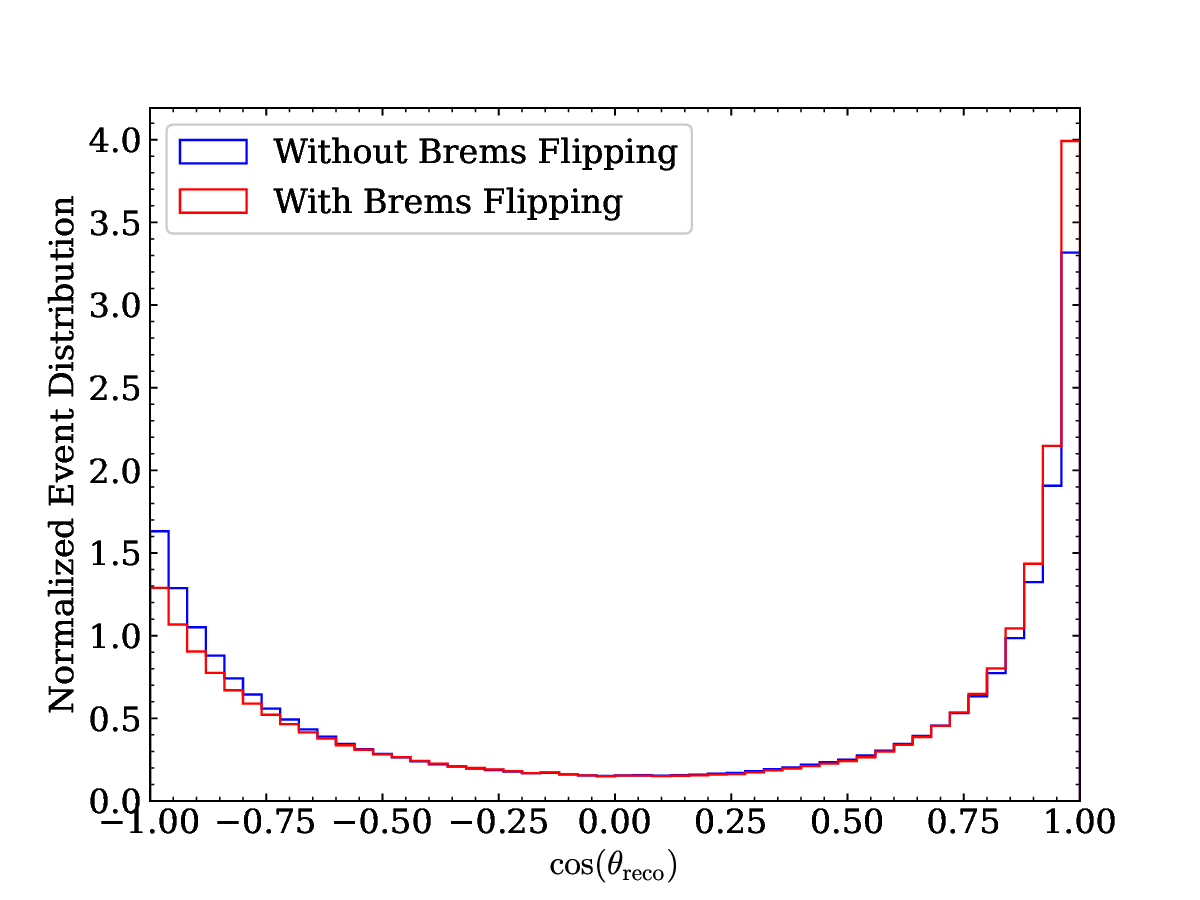}
    \includegraphics[width=\columnwidth]{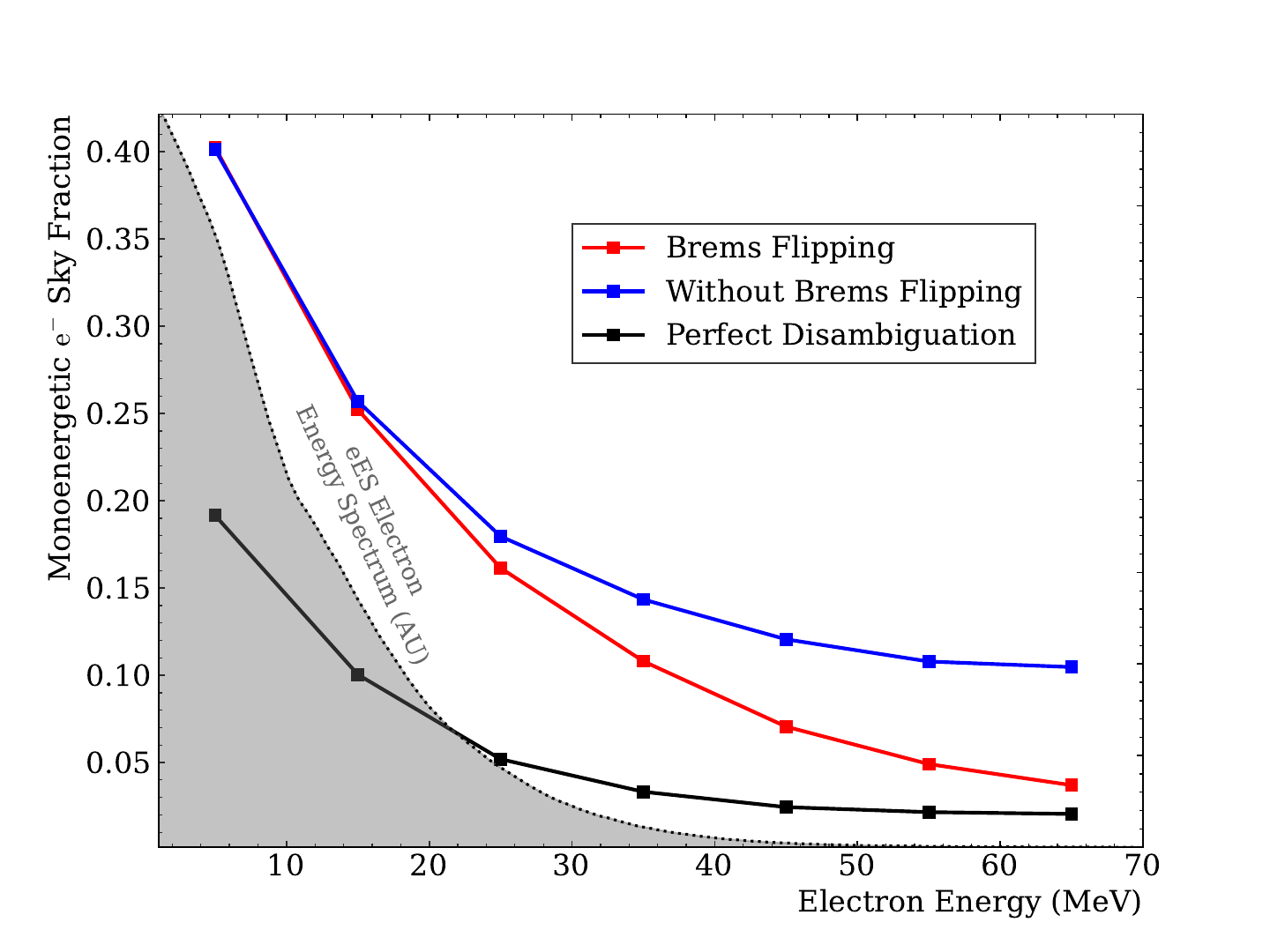}
    \caption{Effectiveness of the brems-flipping algorithm: The top plot shows the bimodal distribution of the angular difference between true and reconstructed electron directions using the supernova eES electron energy spectrum, centering around the parallel ($\cos{\theta} = 1$) and anti-parallel ($\cos{\theta} = -1$) directions. Brems flipping noticeably decreases the magnitude of the anti-parallel peak. The bottom plot shows the relationship between the covered sky fraction and mono-energetic electron energy. The black curve corresponds to a perfect directional disambiguation always resulting the true direction. Brems flipping performs better at higher energies, as there are more secondary tracks to reference. The electron energy spectrum from eES in gray illustrates the energies relevant to supernova pointing.}
    \label{fig:daughter_flipping}
\end{figure}

The performance of the reconstruction algorithm before the application of brems flipping is shown in the top plot of Fig.~\ref{fig:daughter_flipping} in blue. The distribution is centered around both the parallel and anti-parallel directions with respect to the true direction, with a minor preference for the parallel direction. The reason for this preference is that the projection matching algorithm creates many track candidates at first and then merges candidates that are close to each other and have a small angle between them. If a secondary track is sufficiently close to the primary track, the merging process may combine it with the primary track, flipping the primary track to the correct orientation. 
Brems flipping extends the same principle to the separately detected brems particles, meaning that their location with respect to the track indicates the direction of the primary track.
By applying brems flipping, the magnitude of the distribution in red in the anti-parallel direction is noticeably decreased, confirming that the technique is a valuable tool in resolving the ambiguity in track directions.

The bottom plot of Fig.~\ref{fig:daughter_flipping} shows how brems flipping reduces the sky fraction for mono-energetic electron energies. The electron energy spectrum from eES events is given as a reference for the energy regime of interest for supernova pointing. Brems flipping has a higher impact at higher primary electron energies, as those result in more secondary tracks. This fact motivates us to include energy reconstruction as part of the supernova direction reconstruction, as much can be gained by weighing higher-energy events more than lower-energy ones. The successful application of brems flipping on the full supernova neutrino MC is presented in Sec.~\ref{sec5}.

\begin{figure}[h!]
    \centering
    \subfigure[]{\includegraphics[width=\columnwidth]{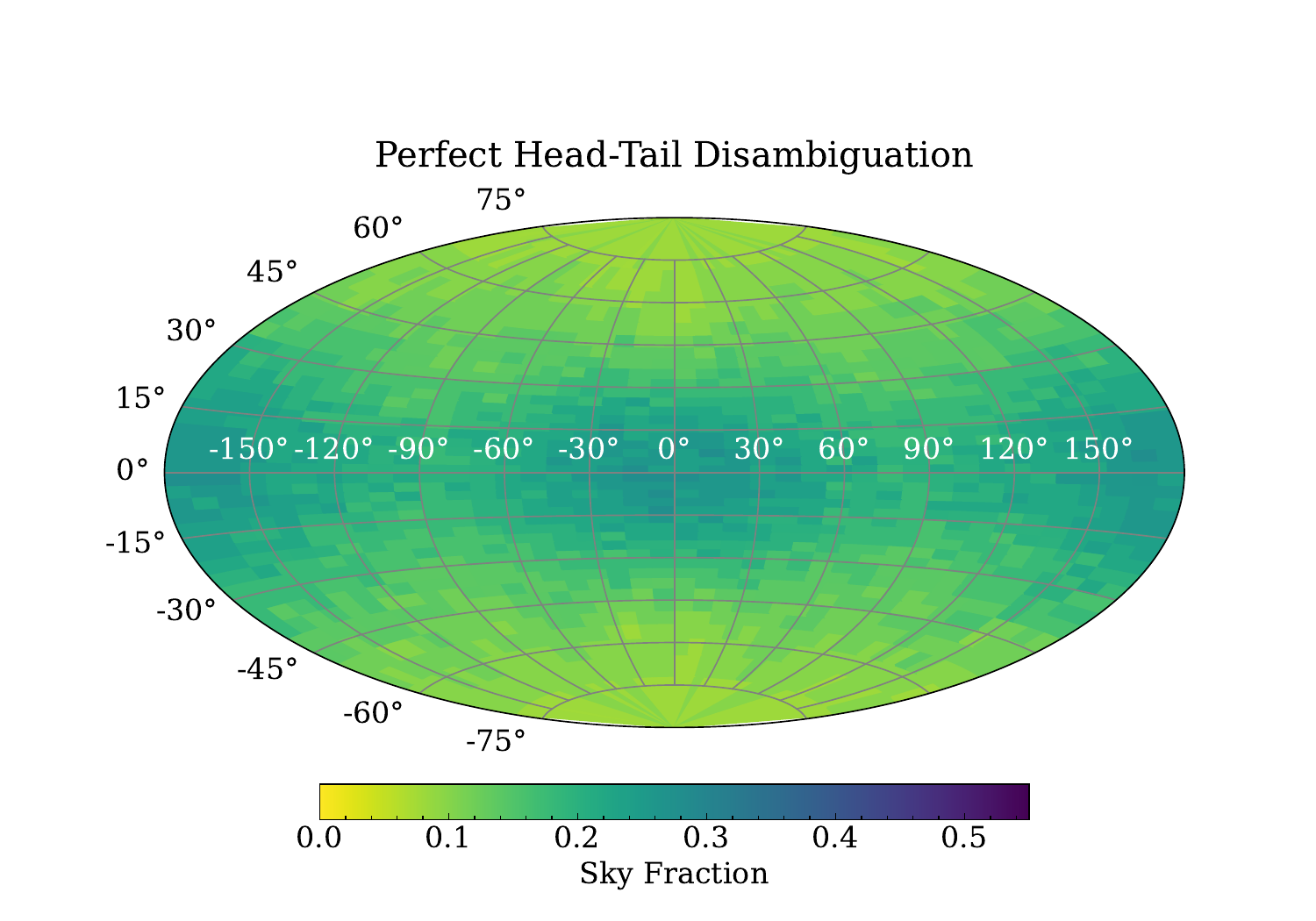}}
    \subfigure[]{\includegraphics[width=\columnwidth]{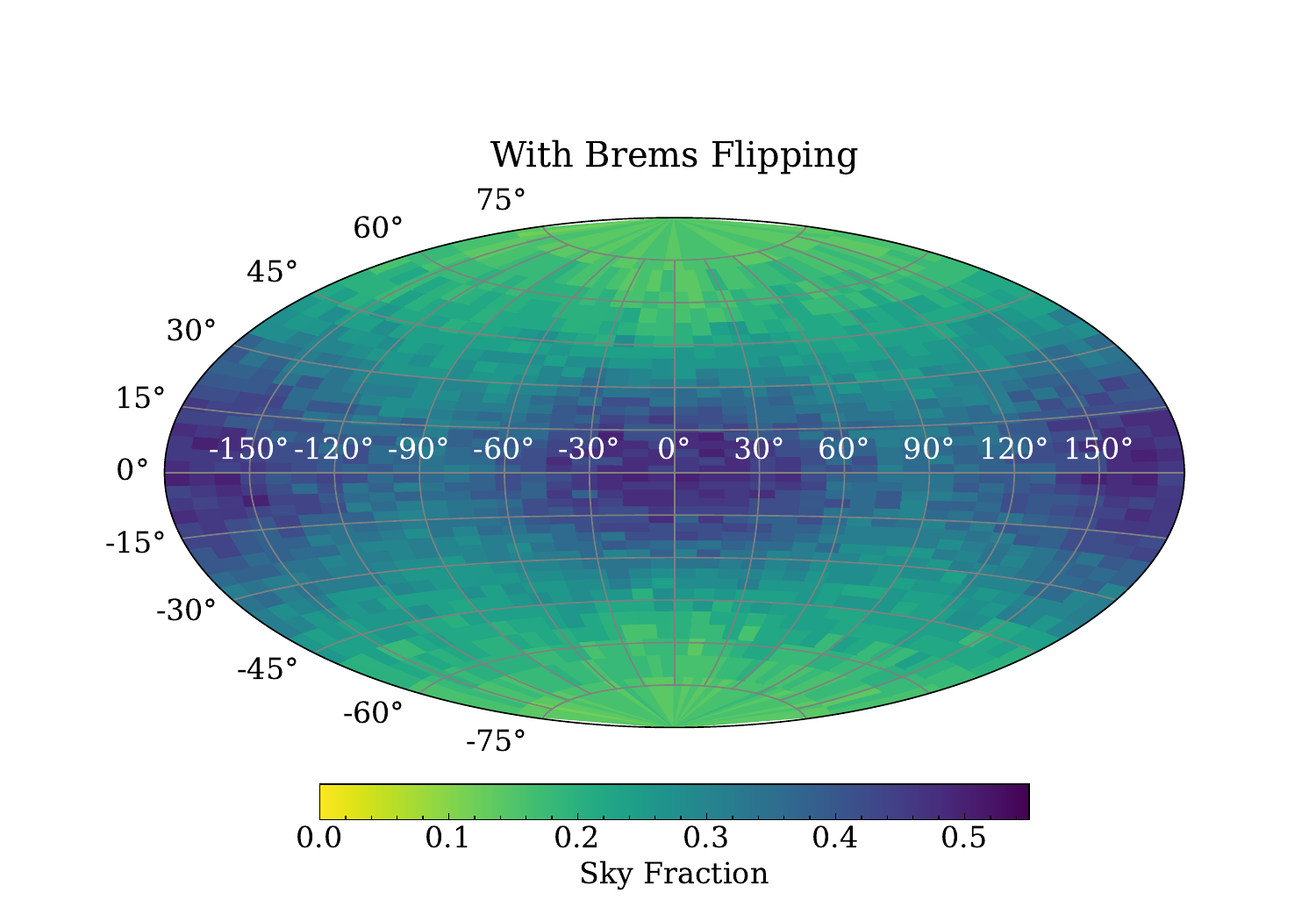}}
    \caption{Pointing resolution, defined in terms of ``sky fraction" as described in the text, of single electron events as a function of the electron's true direction in the detector coordinate system to study anisotropies in the performance. The coordinate system is defined with $\pm\hat{x}$ being the drift directions, and $+\hat{z}$ being approximately the beam direction. $\theta$ (shown vertically) and $\phi$ (shown horizontally) are spherical coordinates in a coordinate system for which $\theta = 90^\circ$ correspond to the $+\hat{z}$ direction and $\theta = 0^\circ, \phi=0^\circ$ corresponds to the $+\hat{x}$ direction. The supernova eES electron energy distribution is used. 
    (a) shows the pointing resolution given perfect track head-tail disambiguation (b) shows the pointing resolution for the actual performance of the reconstruction algorithm including the brems-flipping algorithm.}
    \label{fig:electron_anisotropy}
\end{figure}

Our studies show that the performance of the tracking algorithm as a function of the initial track direction with respect to the detector coordinate system is not perfectly isotropic, as expected. The particle track's inclination with respect to the readout wire planes affects the charge deposition on the wires~\cite{Antonello:2012hu} and therefore influences the uncertainty of the reconstructed direction. The worst performance is observed along the drift direction (detector coordinate $\pm\hat{x}$) due to projection of the track over fewer wires.
Figure~\ref{fig:electron_anisotropy}(a) demonstrates this anisotropy. For this figure, it is assumed that there are no directional ambiguities, to reveal the full impact of this anisotropy.
Figure~\ref{fig:electron_anisotropy}(b) is created assuming the actual performance of the reconstruction algorithm including brems flipping for the head-tail disambiguation.
While the reconstruction performance varies with track direction, the figures show that the track direction ambiguity has a stronger detector-related anisotropy. We therefore first evaluate pointing capability assuming a uniform performance in directional reconstruction. Subsequently, we investigate the performance of the method as a function of supernova direction in detector coordinates. 
 
\section{Maximum likelihood method for burst pointing}
\label{sec5}
A maximum likelihood method~\cite{abe2016real} is used to reconstruct the direction of a supernova burst ensemble of events from the reconstructed information of individual neutrino events. For an event with a known interaction type, the probability density function (PDF) has the functional form of $p_{r}(E_i, \hat{d}_i; \hat{d}_{SN})$, where the index $r$ indicates the neutrino channel: eES or $\nu_e$CC. $E_i$ and $\hat{d}_i$ are the reconstructed energy and direction of the specific event, and $\hat{d}_{SN}$ is the direction of the supernova. The PDF is formulated to be a function of the reconstructed energy as well as the inner product of the supernova direction and reconstructed electron direction, $\hat{d}_i \cdot  \hat{d}_{SN} = \cos{\theta_{SN,i}}$, as we make the aforementioned assumption that the event direction reconstruction quality is uniform as a function of the true neutrino direction\footnote{In principle, residual detector response anisotropy can be taken into account in the likelihood PDFs.}. The PDFs are normalized by energy bin, as the electron energy distribution does not affect the direction likelihood. \larsoft{} simulations of one million events in each channel are used to determine the PDFs. The Monte Carlo samples are divided into energy bins, spanning from 0 to 40\,MeV with a width of 2\,MeV per bin, from 40\,MeV to 70\,MeV at 10\,MeV per bin, and a final bin spanning from 70\,MeV from 100\,MeV. Varying energy bin widths are used to ensure sufficient counting statistics at high energies, where the expected numbers of events are low.

\begin{figure}[h!]
    \centering
    \includegraphics[width=\columnwidth]{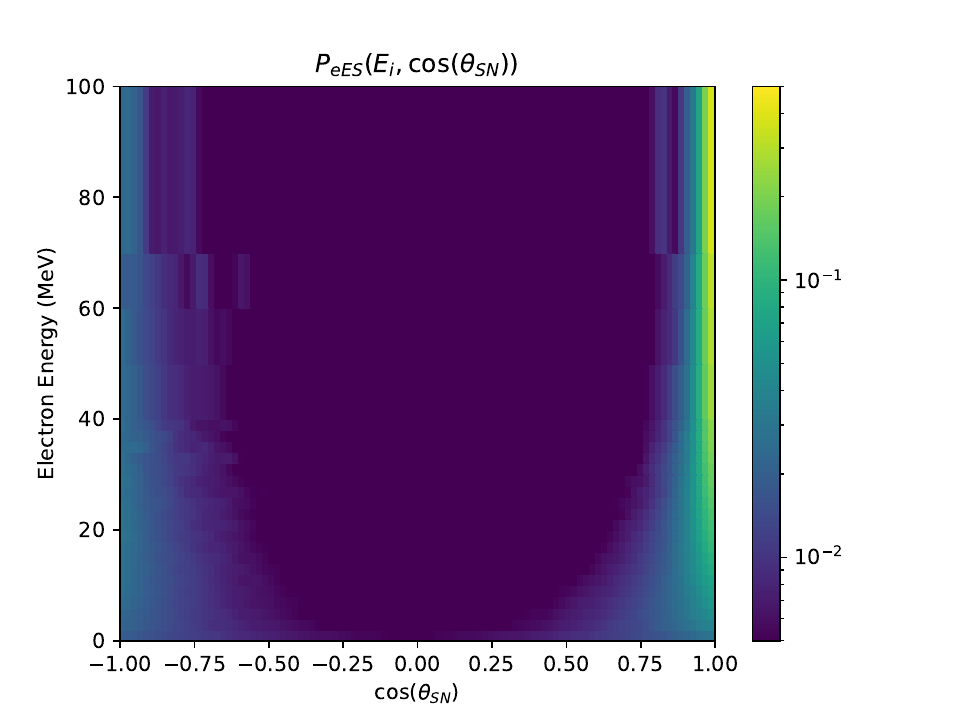}
    \includegraphics[width=\columnwidth]{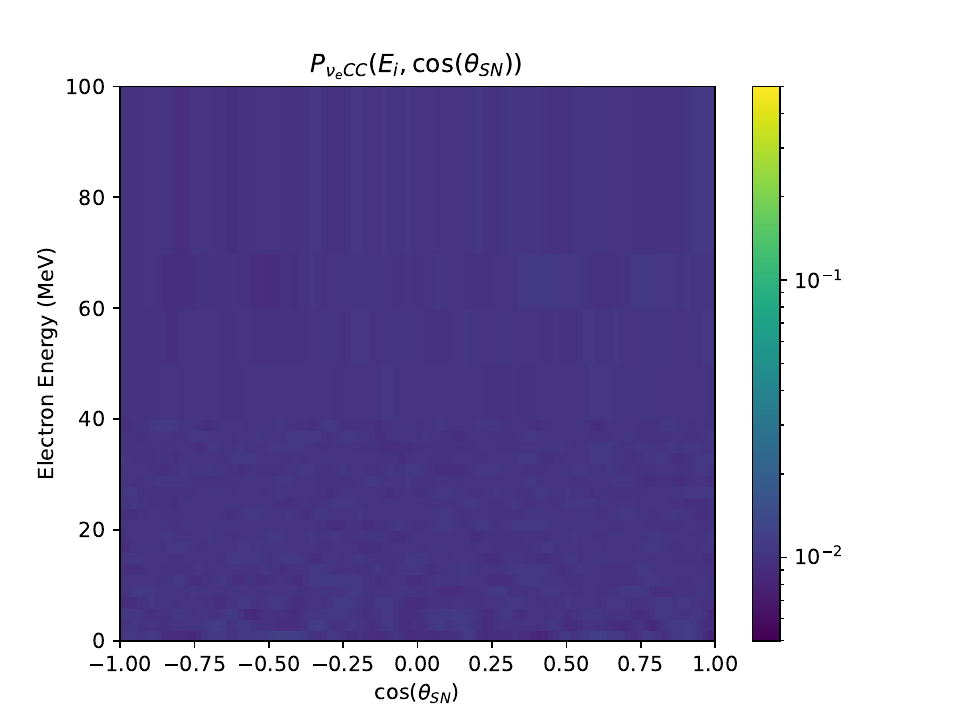}
    \caption{PDFs of eES and $\nu_e$CC events. The eES PDF demonstrates a high correlation between the supernova and primary electrons' direction increasing towards higher electron energies. The $\nu_e$CC PDF is mostly flat, as the correlation between primary electron and neutrino directions is weak.}
    \label{fig:2dpdfs}
\end{figure}

\begin{figure}[h!]
    \centering
    \includegraphics[width=\columnwidth]{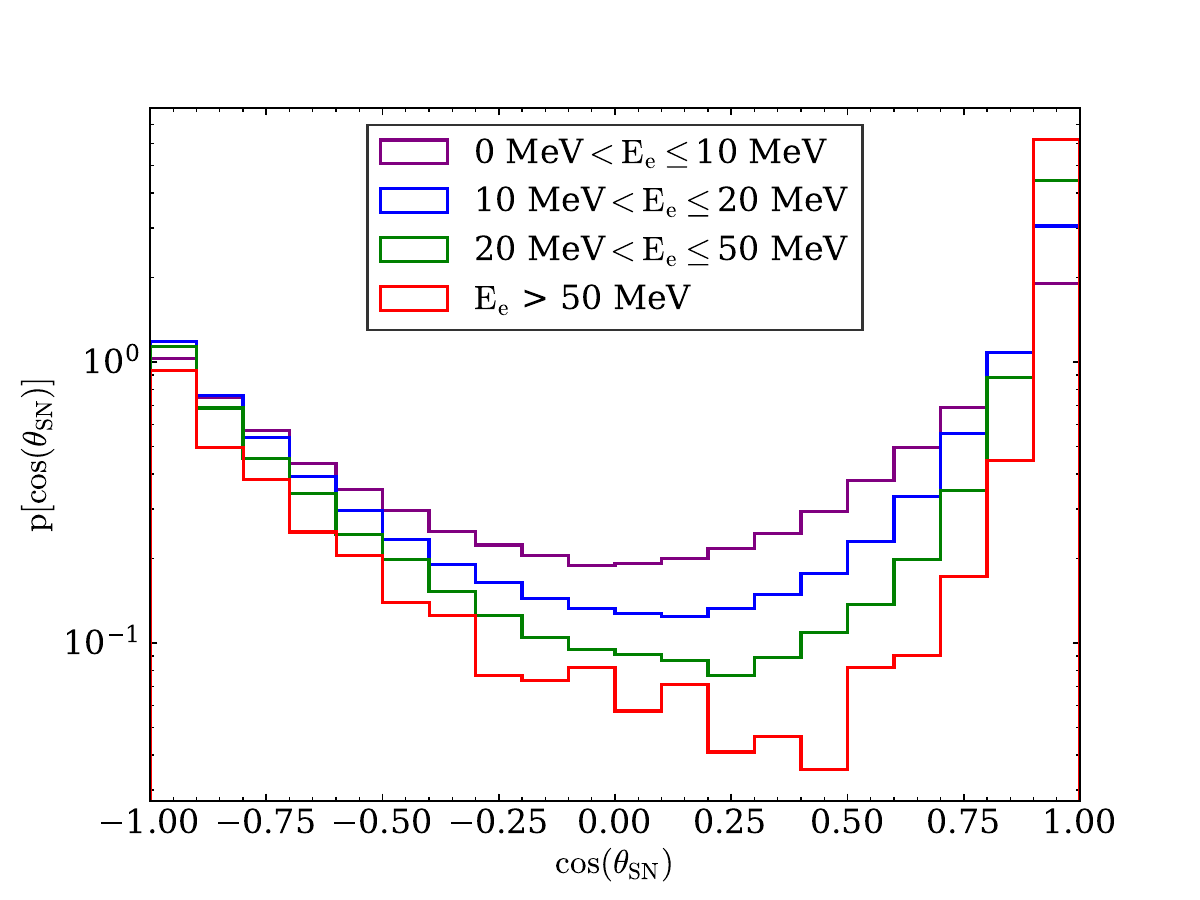}
    \caption{Several bins for the PDF of the eES interaction in Figure \ref{fig:2dpdfs}, re-binned to 10\,MeV / bin. The relative magnitude of the peak near $\cos{\theta_{SN}} = 1$ (the parallel direction) increases as energy increases because brems flipping improves at high energies. The width of the true direction peaks also decreases, indicating a decrease in directional variance.}
    \label{fig:pdf_slices}
\end{figure}

The generated PDFs are shown in Fig.~\ref{fig:2dpdfs} for both eES and $\nu_e$CC events. For the $\nu_e$CC channel, the low correlation between electron and neutrino directions renders the PDF mostly flat. The eES interactions demonstrate a high correlation between the supernova direction and the reconstructed direction. A bimodal distribution is seen due to the tracks' head-tail ambiguity, but the peak for the true direction (centered around $\cos{\theta_{SN}} = 1$) has a much higher amplitude than the peak around the flipped direction ($\cos{\theta_{SN}} = -1$) thanks to brems flipping. In particular, at higher electron energies, the pointing of eES events improves both in variance around the supernova direction, as well as in track directional disambiguation (the same effect as can be observed in Figure \ref{fig:daughter_flipping}). The improved performance at high energy is further demonstrated in projections of this PDF for various energy ranges shown in Figure \ref{fig:pdf_slices}.

\begin{figure*}[htb]
    \includegraphics[width=0.8\textwidth]{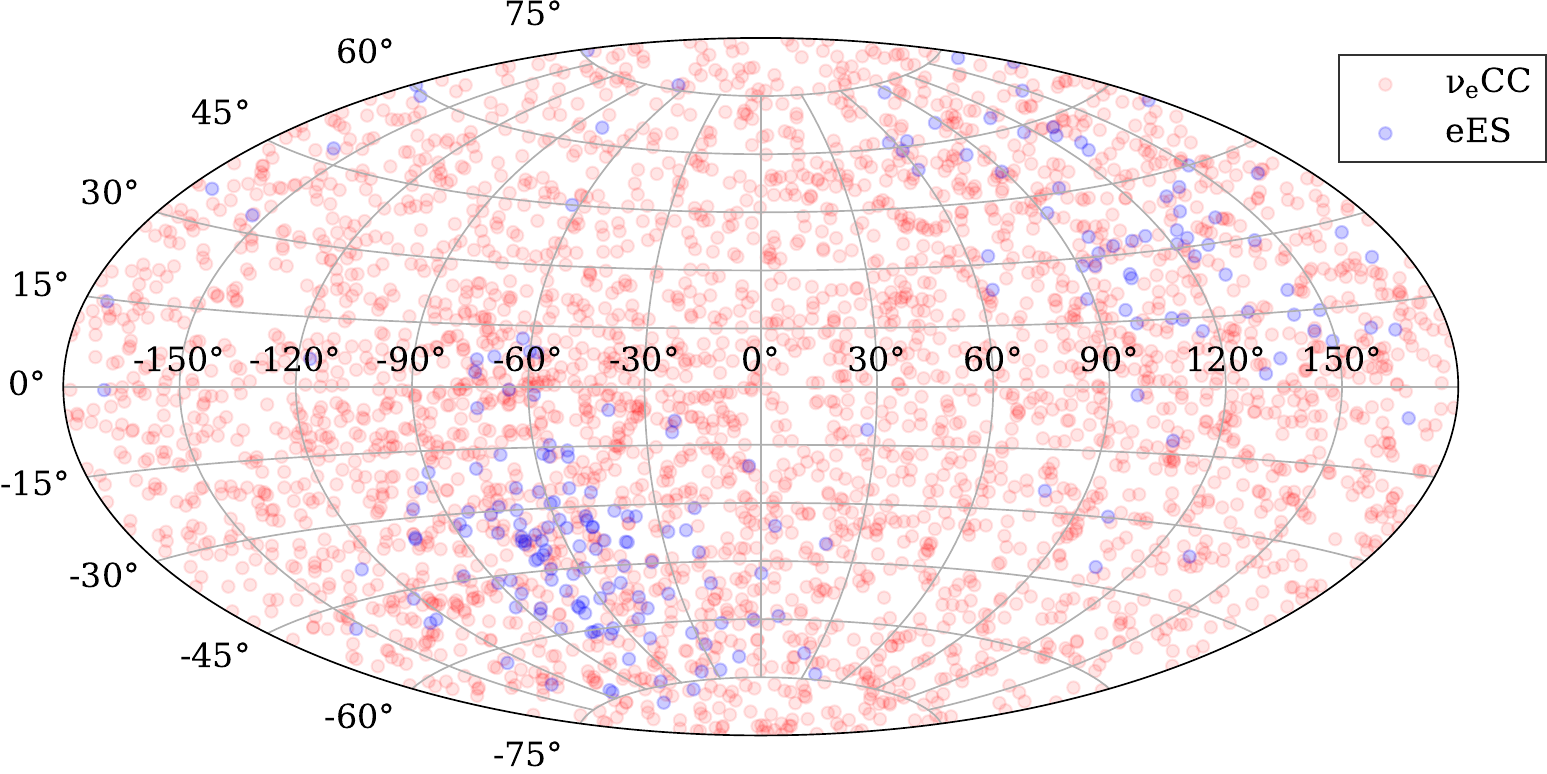}
    \caption{An example directional map filled with the reconstructed electron directions for a simulated supernova burst, eES events carrying the directional information are marked in blue. Event statistics are shown for a core collapse at a distance of 10\,kpc, and 40\,kton of fiducial mass in the detector.}
    \label{fig:burst_events}
\end{figure*}

\begin{figure*}[htb]
    \centering
    \includegraphics[width=0.8\textwidth]{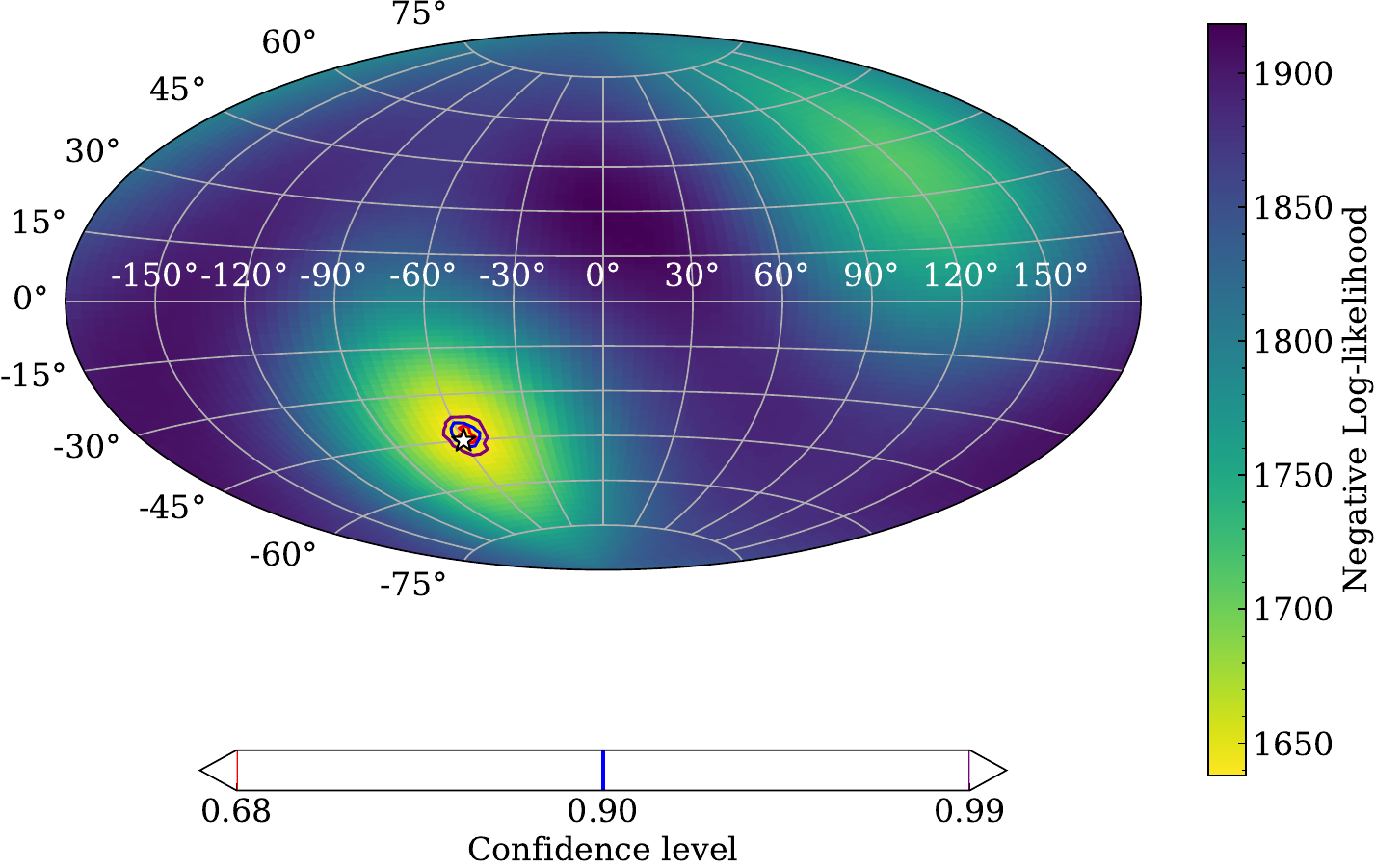}
    \caption{An example directional map filled with the negative log-likelihood values and confidence contours. This map is computed from the same burst shown as in Fig.~\ref{fig:burst_events}. Reconstruction is done assuming the classification parameters of $c_{eES\rightarrow eES} = 0.86$ and $c_{\nu_eCC\rightarrow eES} = 0.04$ and a successful direction reconstruction is achieved with the actual supernova direction marked with a star in the figure.}
    \label{fig:skymap}
\end{figure*}

The supernova direction can be reconstructed from the PDFs with the maximum likelihood method. Given a set of reconstructed events, the log-likelihood as a function of supernova direction is expressed as:
 \begin{equation}
    \begin{split}
        - \log \mathcal{L}(\hat{d}_{SN}) &=- \sum_i \log p(E_i, \hat{d}_i; \hat{d}_{SN})
    \end{split}
    \label{eq:neglog}
\end{equation}
Here, a perfect classification between eES and $\nu_e$CC events is assumed, $p = p_{eES}$. The effect of imperfect classifications is discussed later. By parameterizing $\hat{d}_{SN}$ using the azimuth and zenith angles, the supernova direction is reconstructed via a two-variable minimization. For the minimization, events that have reconstructed electron energies of less than 5\,MeV are excluded, as they are found to degrade pointing performance due to the relatively large uncertainties in their energy and direction reconstruction. The minimization is done by an adaptive grid search: after the initial coarse search of the entire phase space we applay a second grid search of a smaller area. The result of the second grid search is further refined using a local minimization of the PDF initialized at the fitted grid point.
The reconstructed neutrino directions associated with a typical supernova burst are shown in Fig.~\ref{fig:burst_events}, and the negative-log likelihoods values for all possible supernova directions for this burst are shown in Fig.~\ref{fig:skymap}. Two local minima are 180\,degrees away from each other due to the remaining ambiguity in track direction.

\section{Performance of reconstruction algorithm on supernova simulations}
\label{sec6}
\begin{figure}[htb]
    \centering
    \includegraphics[width=\columnwidth]{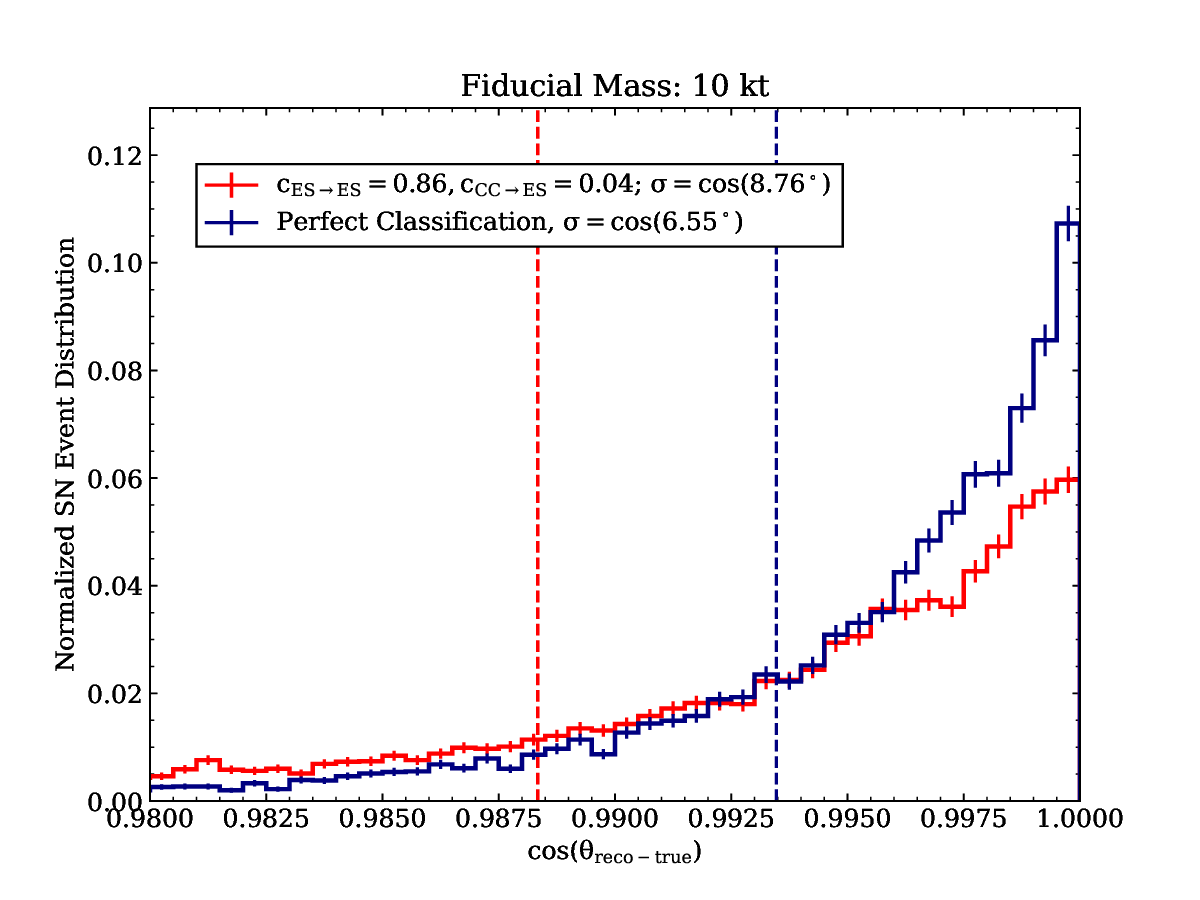}
    \includegraphics[width=\columnwidth]{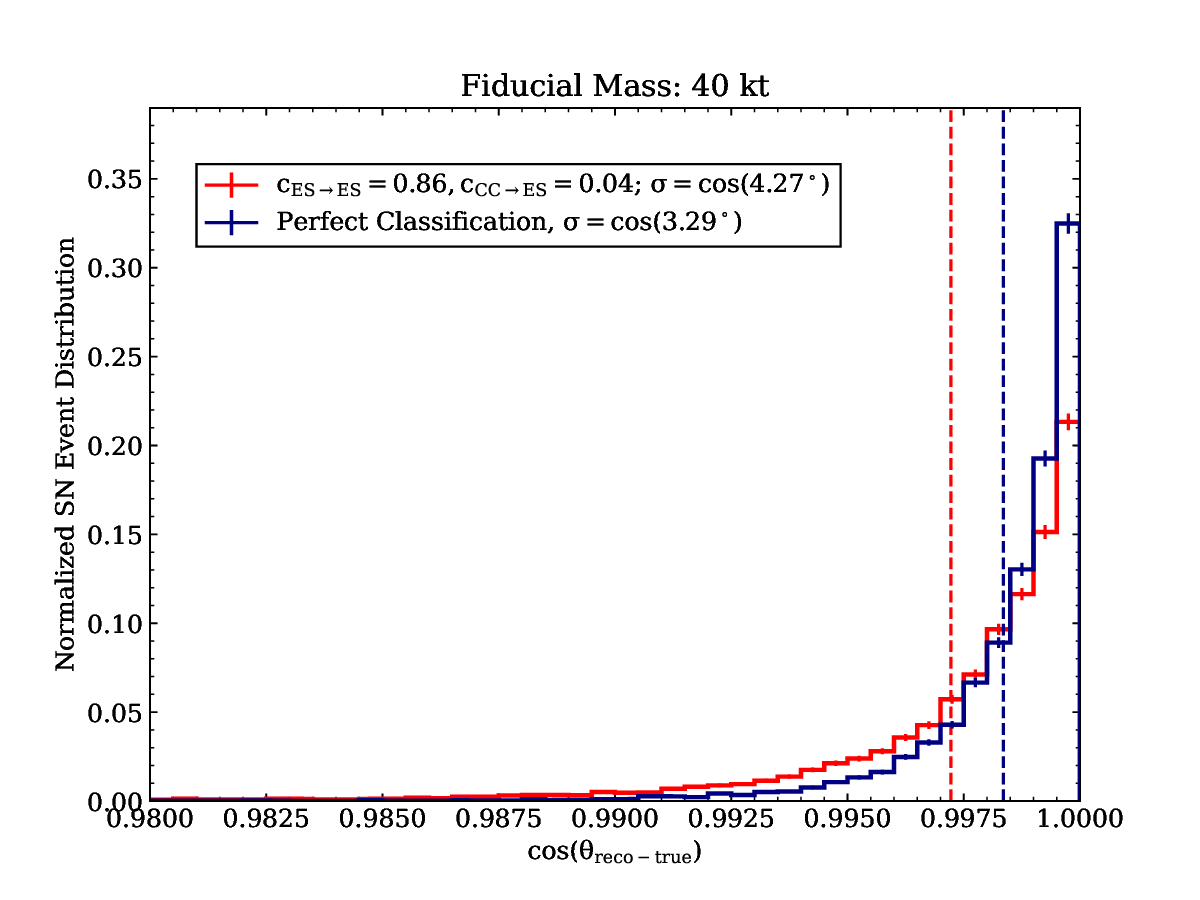}
    \caption{Distribution of the angular difference between the reconstructed and true supernova direction, for 10,000 simulated supernova bursts. The distribution is shown for both perfect event classification and for an assumed 4\% misclassification of $\nu_e$CC events as eES as described in \cite{erin_conley_2020_4122909}. The ranges to the right of the respective colored dashed lines correspond to the 68\% confidence intervals. The few bursts with a flipped reconstructed direction are excluded from the figure. The top figure corresponds to 10\,kton fiducial mass, while it is 40\,kton for the bottom figure.}
    \label{fig:pointres}
\end{figure}

Supernova bursts are simulated to evaluate the performance of the reconstruction algorithm. The simulation is carried out by randomly picking eES and $\nu_e$CC events from a large pool of events with uniformly distributed neutrino directions. The number of events selected per burst is based on the expected number of interactions for a 10\,kpc supernova, as calculated by \snowglobes~(see Tab.~\ref{tab:event_rates}). For each selected event, the reconstructed direction is rotated in such a way that the true burst directions align for all events. To account for the effect of the anisotropic direction reconstruction performance, events are only selected if their neutrino direction lies within a cone of a 10\,degree opening angle of the randomly selected supernova direction. In total, 10,000 supernova bursts are simulated. Assuming no mis-identification of event types and using only the reconstrucred eES events, and a pointing resolution (as defined in the previous section) of 3.4\,degrees for a fiducial mass of 40\,kton (four modules, full setup) and 6.6\,degrees for one module of 10\,kton is achieved. 
The distribution of the truth-to-reconstruction angular difference for the burst is shown in Fig.~\ref{fig:pointres} for two different assumptions on event classification quality. It is worth noting that only $\sim$0.1\% of the simulated bursts' reconstructed directions differed more than 90\,degrees from the supernova position thanks to the brems flipping for both classification cases. The reconstructed directions of poorly reconstructed bursts are very close to being anti-parallel to the supernova direction.

In reality, the capability to distinguish between eES and $\nu_e$CC events, with the latter carrying poor pointing information, needs to be considered. The quality of a realistic event classifier can be quantified in the form of a confusion matrix:
\begin{equation}
    C = \begin{bmatrix}c_{eES\rightarrow eES} & c_{eES\rightarrow \nu_eCC}\\c_{\nu_eCC\rightarrow eES} & c_{\nu_eCC\rightarrow \nu_eCC}\end{bmatrix}
\end{equation}

The elements of this matrix are expressed in the form of $c_{A\rightarrow B}$, which describes the portion of events of interaction type $A$ that are classified as interaction type $B$. The matrix is normalized to be independent of the expected number of events for each interaction. Assuming no loss of events due to detector inefficiency, the rows of the matrix will sum to one. The worst-case scenario would correspond to a confusion matrix with all of its elements being $1/2$, which effectively represents a completely random grouping of all events into two groups.

\begin{figure}[h]
    \centering
    \includegraphics[width=\columnwidth]{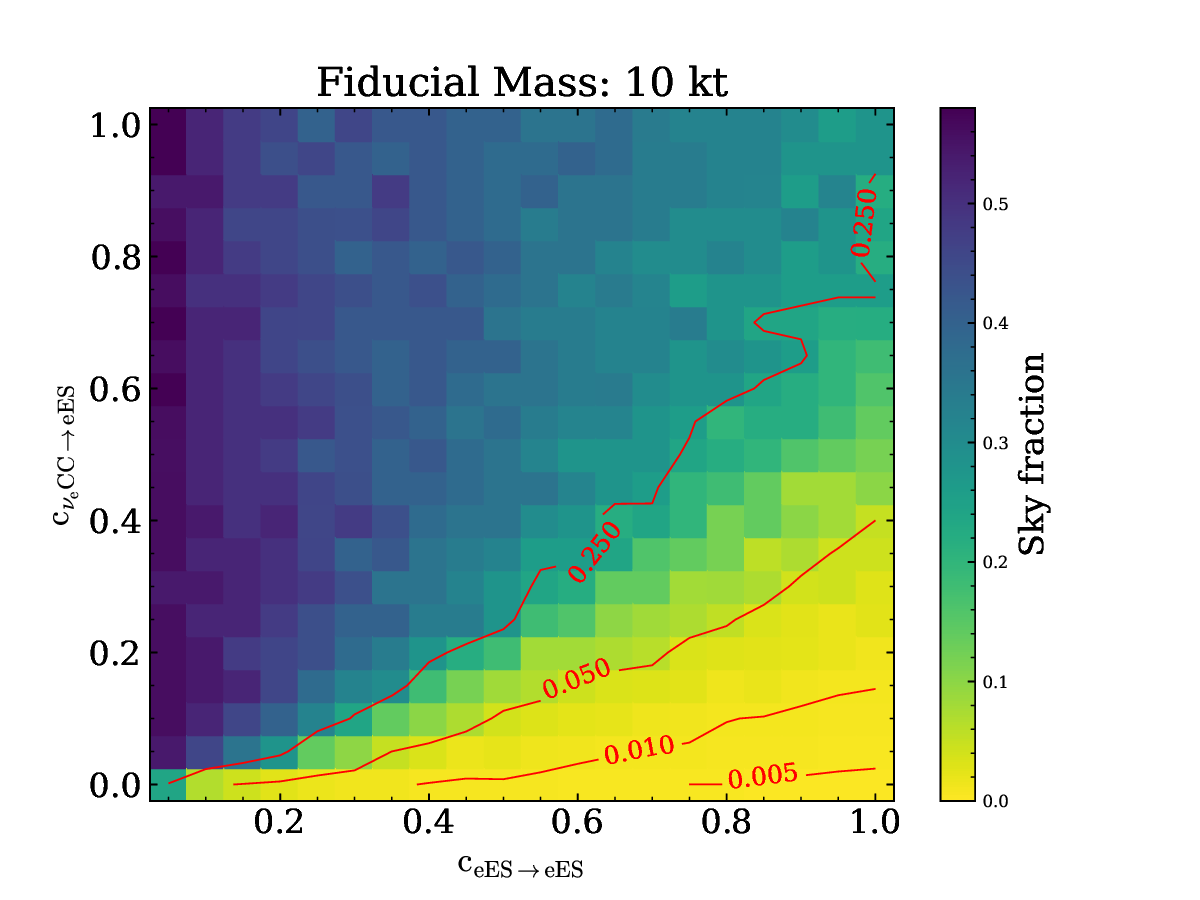}
    \includegraphics[width=\columnwidth]{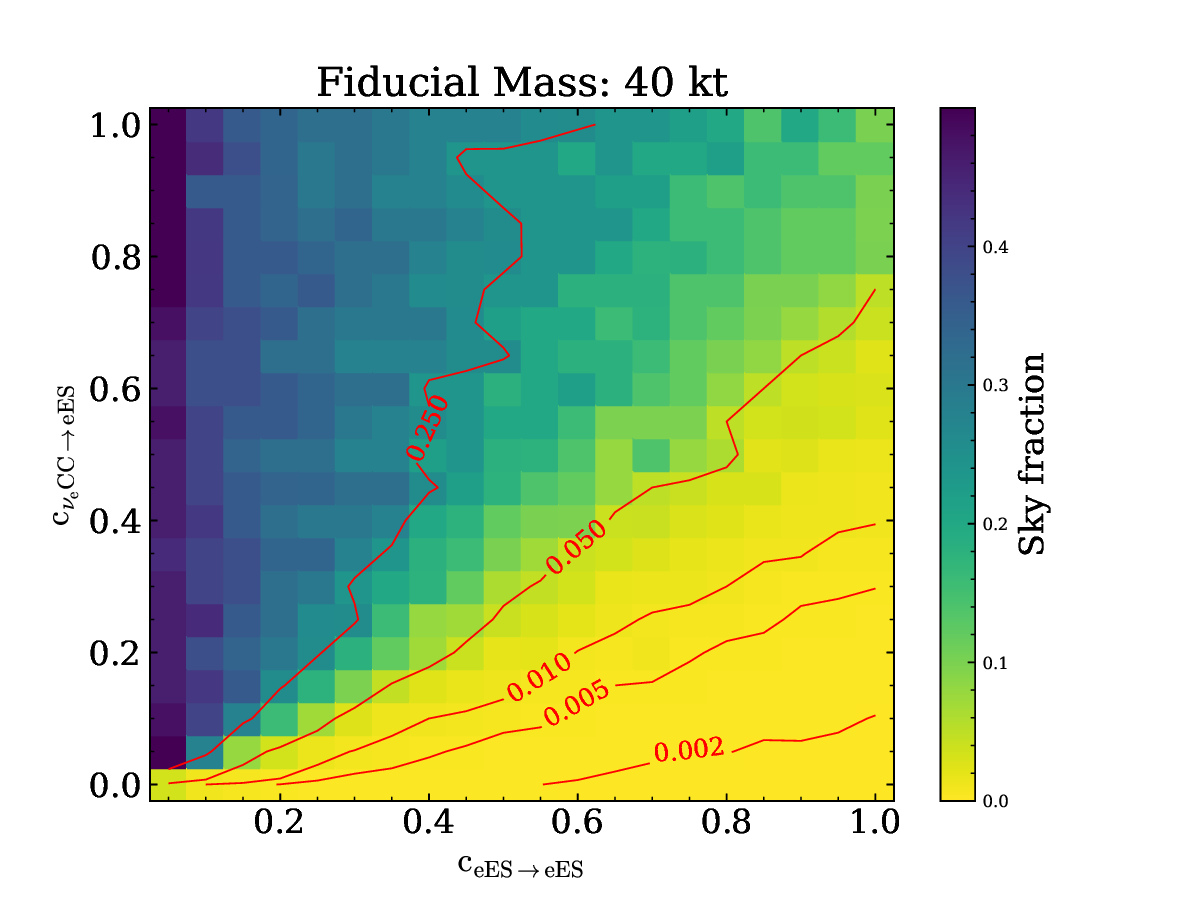}
    \caption{Burst pointing resolution as a function of eES true positives ($c_{eES\rightarrow eES}$) and $\nu_e$CC false negatives ($c_{\nu_eCC\rightarrow eES}$). Results assuming the fiducial mass of a single far detector module (10\,kton) and all four planned modules (40\,kton) are shown. For each pair of values, 1000 supernova bursts are simulated to determine the pointing resolution. Contour lines for various pointing resolution angles are also shown.}
    \label{fig:mixing}
\end{figure}

In adopting a more realistic classifier taking into account the misidentification of events, the aforementioned reconstruction procedure is effectively unchanged, except that the PDF $p$ used in (\ref{eq:neglog}) no longer corresponds to $p_{eES}$, but is rather a sum of $p_{eES}$ and $p_{\nu_eCC}$, weighted by the expected number of events from each interaction in the classification channel, which should be known ahead of time. As pointing information is primarily accessible from eES events, only the eES classification channel is used for direction reconstruction. Subsequently, the only elements in the confusion matrix $C$ that affect the pointing resolution are $c_{eES\rightarrow eES}$ and $c_{\nu_eCC\rightarrow eES}$. Figure~\ref{fig:mixing} shows the pointing resolution as a function of these two matrix elements. For the worst-case scenario (where both elements are equal to $0.5$), the pointing resolution is $102$\,degrees, averaging over a flat distribution of supernova directions. While highly precise classifiers would provide highly accurate pointing results, even a weak classifier can yield a useful pointing resolution. At $c_{eES\rightarrow eES} = 0.6$ and $c_{\nu_eCC\rightarrow eES} = 0.4$, only 20\% better than random selection, the average pointing resolution reaches around 40\,degrees, significant enough to determine the quadrant of the sky that the supernova belongs to. An optimistic estimate of a boosted-decision-tree-based classifier is described in \cite{erin_conley_2020_4122909} at $c_{eES\rightarrow eES} = 0.86$ and $c_{\nu_eCC\rightarrow eES} = 0.04$. With this classifier, the average pointing resolution amounts to $4.3$\,degrees (8.7\,degrees) for a fiducial mass of 40\,kton (10\,kton). The distribution of the reconstruction angle is compared to the perfect classification case in Fig.~\ref{fig:pointres}. While this study provides a rough estimation of the classification capabilities, further validation work is required to understand the classification performance in the presence of noise. Furthermore, because classification performance likely depends on the energy of the event, additional improvements in pointing resolution could be made by leveraging energy regions where the classifier performs best. Moreover, in a future fast online pipeline, real-time channel tagging before the comparison of the reconstructed events to the PDF will likely improve the overall performance as well.

\begin{figure}[h]
    \centering
    \includegraphics[width=\columnwidth]{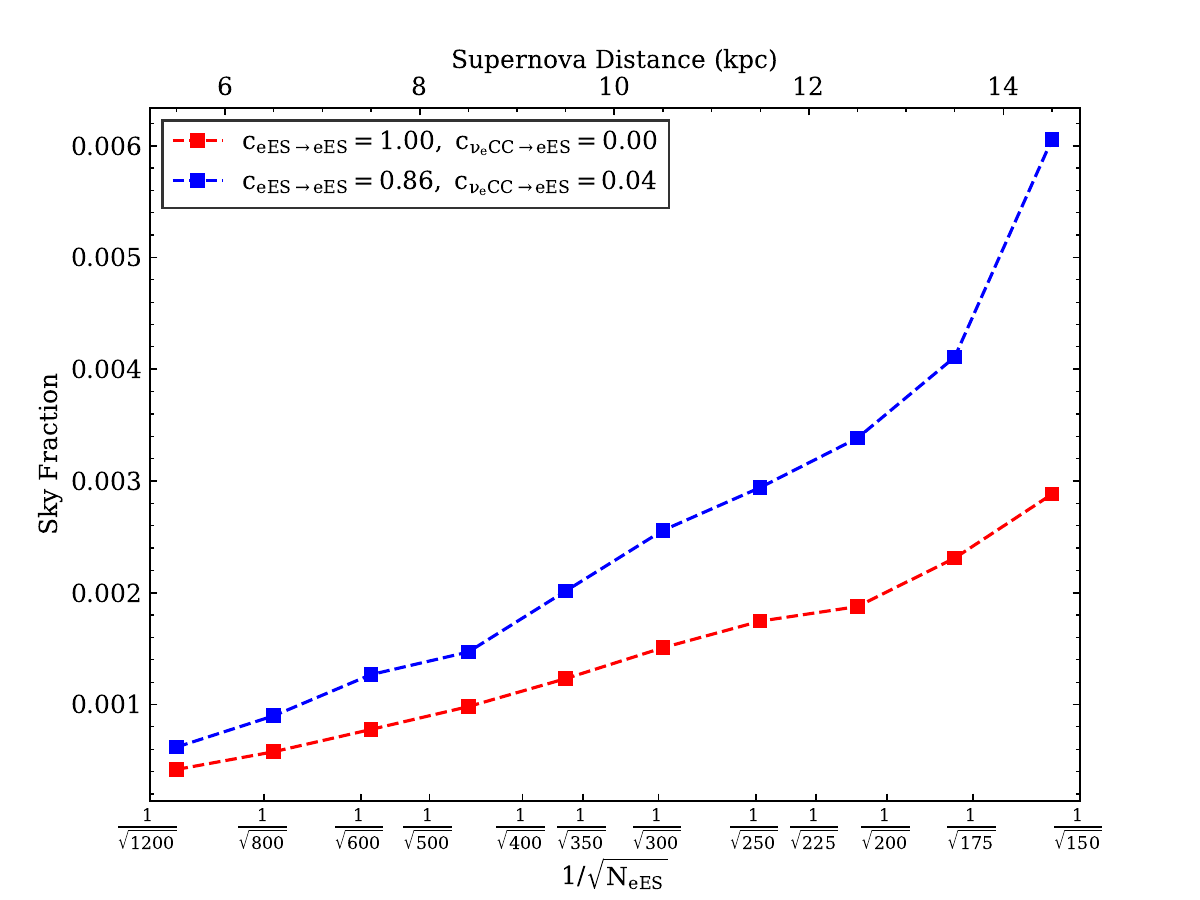}
    \caption{Burst pointing resolution as a function of the number of detected eES events ($N_{eES}$), as well as the corresponding supernova distance. The event rates are calculated assuming the GKVM model at a given distance and are for a fiducial volume of 40\,kton.}
    \label{fig:SNdist}
\end{figure}

The pointing resolution is shown as a function of the number of expected events in Fig.~\ref{fig:SNdist}, along with the corresponding progenitor distance, calculated assuming the GKVM model. The pointing resolution is roughly proportional to the supernova distance, and inversely proportional to the square root of the number of expected events, as expected.

\begin{figure}[h]
    \centering
    \includegraphics[width=\columnwidth]{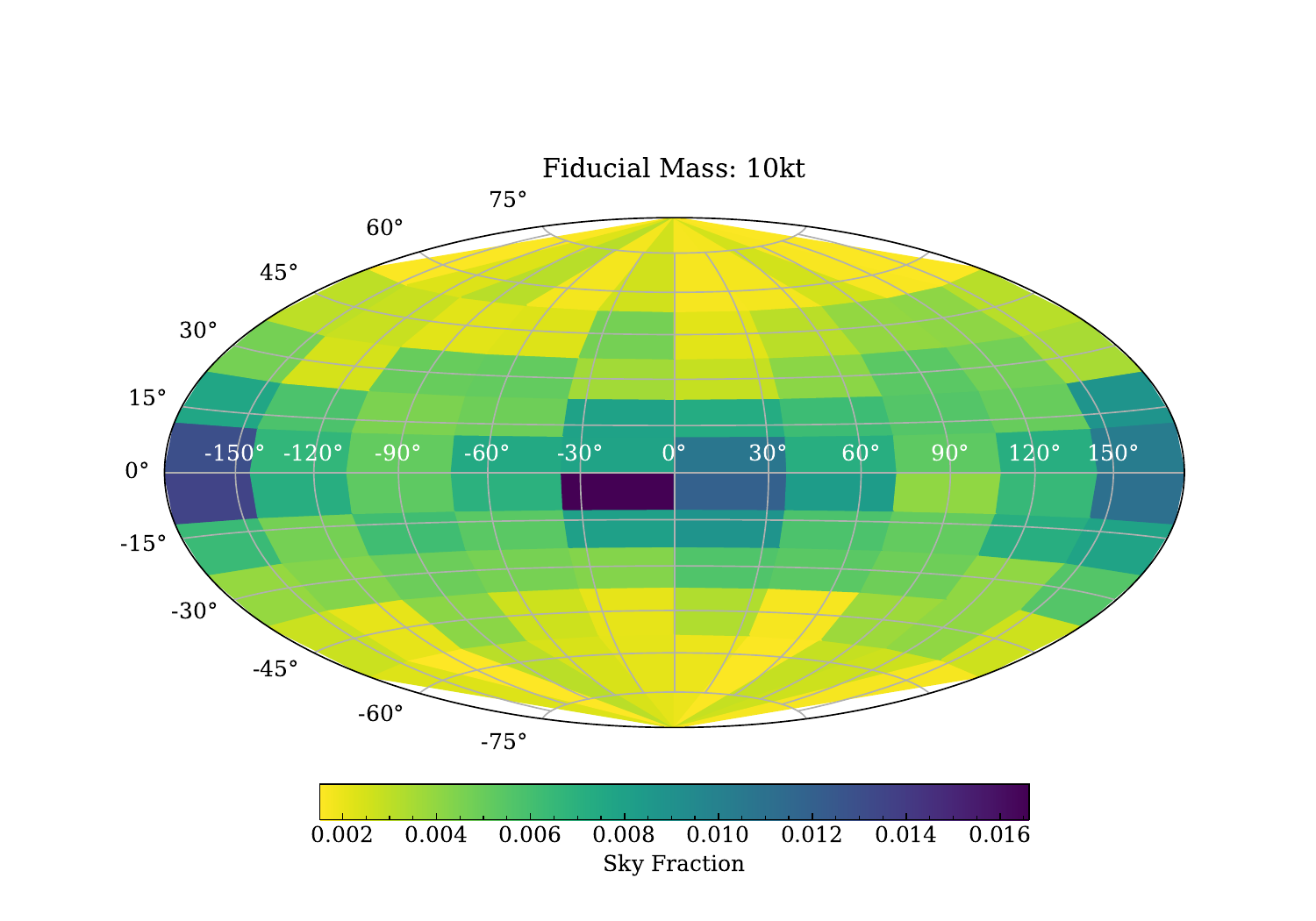}
    \includegraphics[width=\columnwidth]{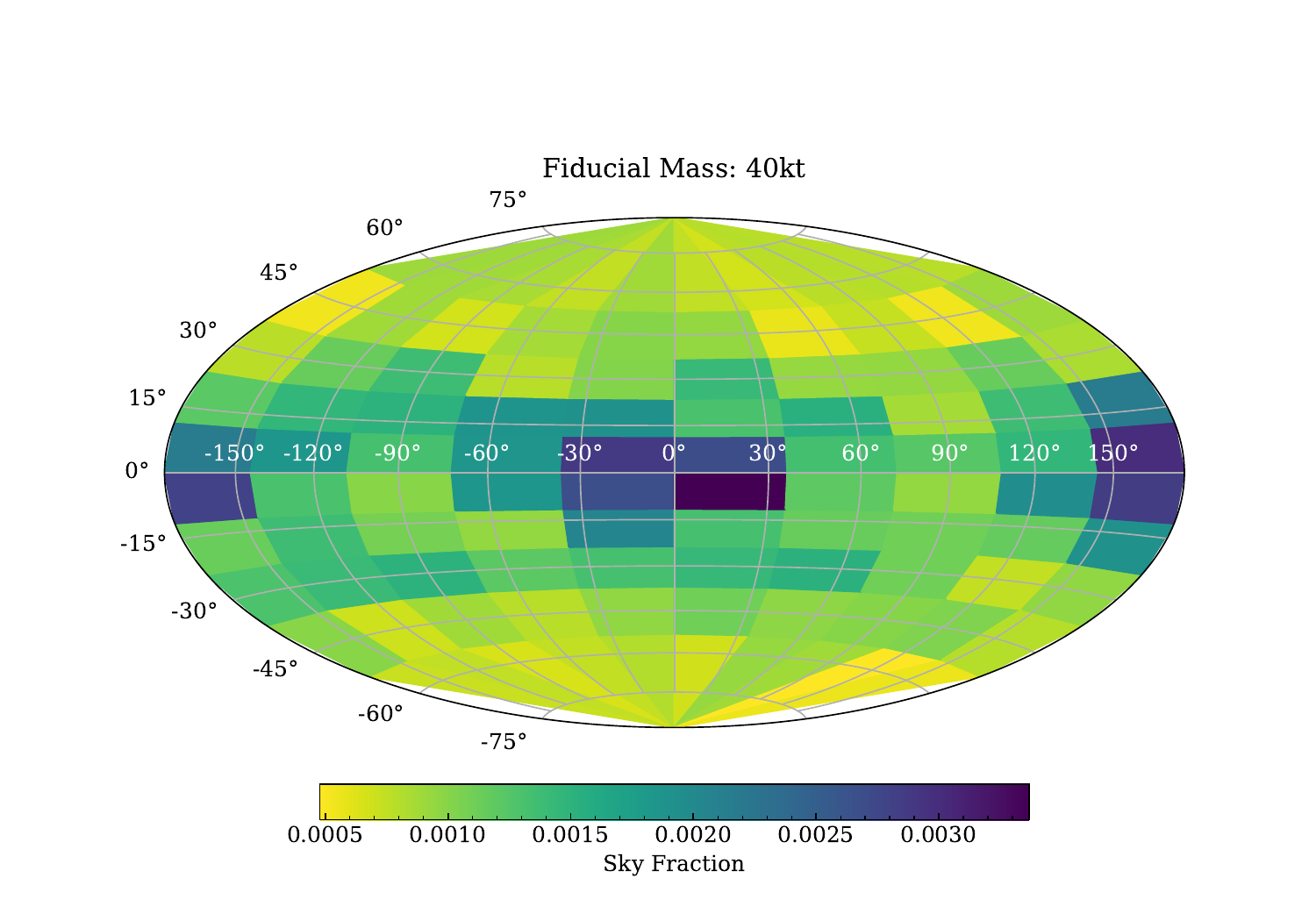}
    \caption{Burst pointing resolution as a function of the direction of the supernova, given in the detector coordinate system ($\pm\hat{x}$ is the drift direction, $+\hat{z}$ is approximately the beam direction). $\theta$ (shown vertically) and $\phi$ (shown horizontally) are spherical coordinates in a coordinate system for which $\theta = 90^\circ$ corresponds to the $+\hat{z}$ direction and $\theta = 0^\circ, \phi=0^\circ$ correspond to the $+\hat{x}$ direction. Pointing resolution is given for the fiducial volumes of 10\,kton and 40\,kton. Perfect classification between eES and $\nu_e$CC events is assumed in these plots. Note that the local pointing resolution is smeared by the event selection radius of 10\,degrees.}
    \label{fig:anisotropy}
\end{figure}

\begin{figure}[h]
    \centering
    \includegraphics[width=\columnwidth]{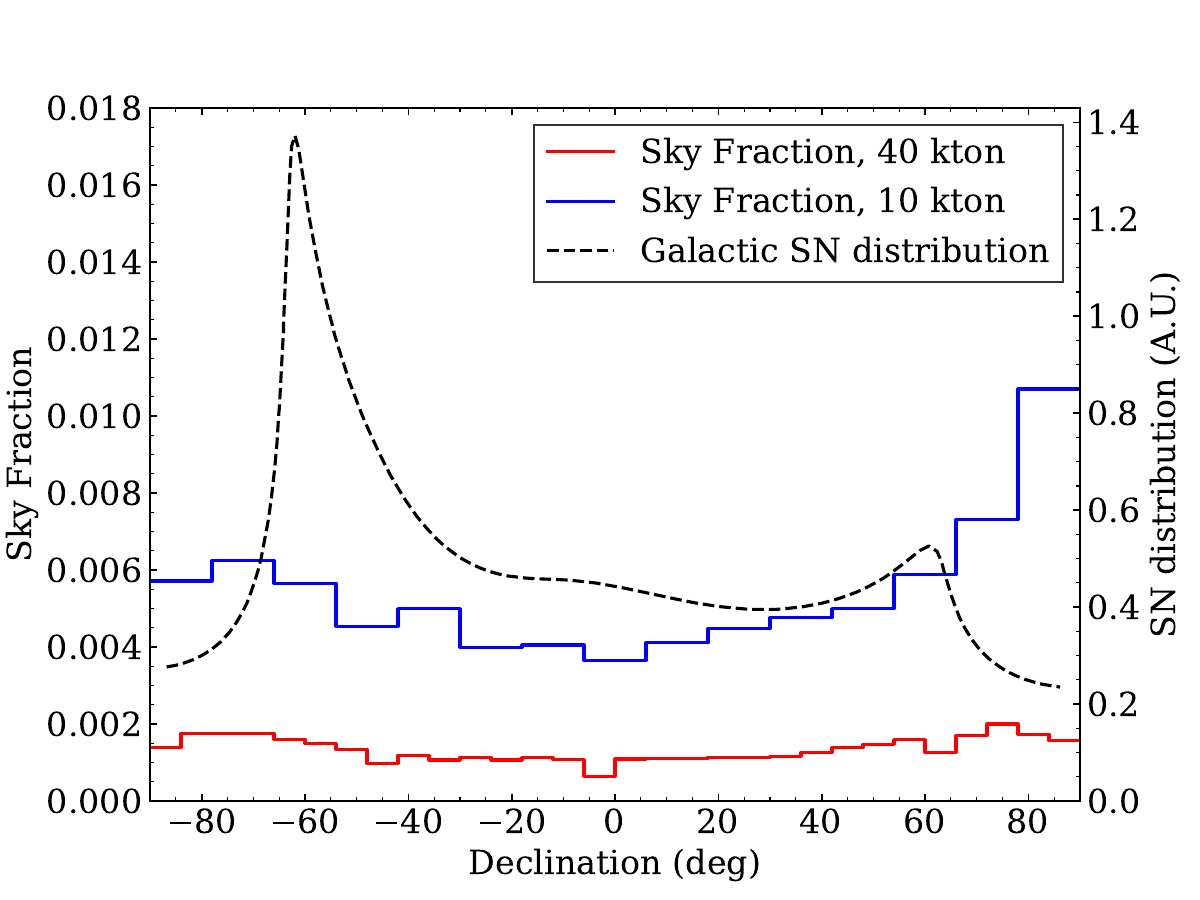}
    \caption{Burst pointing resolution as a function of the supernova direction, shown as a function of declination. The pointing resolution is averaged over right ascension. Shown on a separate axis is the expected declination distribution for galactic supernovae~\cite{mirizzi2006earth}.}
    \label{fig:declination}
\end{figure}

The supernova pointing resolution as a function of the supernova position in the sky is also examined. Fig.~\ref{fig:anisotropy} shows the pointing resolution as a function of the detector coordinates.  As for the results shown in Fig.~\ref{fig:electron_anisotropy}, better pointing resolution can be seen closer to the $\pm\hat{z}$ directions (with $+\hat{z}$ being approximately the beam direction). Note that due to the way supernova bursts are simulated (as described at the beginning of this section), the dependence of the pointing resolution estimate is smeared around the incoming neutrino direction by the event selection radius of 10\,degrees -- but qualitatively, the variation in pointing resolution is of the order of a few degrees. 
Fig.~\ref{fig:declination} shows the pointing resolution in the equatorial celestial coordinate system (RA/Dec), as a function of declination, averaged over right ascension. The figure also depicts the expected declination distribution for galactic supernovae to illustrate the most likely directions for a supernova to occur. Notably, because the $\hat{z}$ direction of the detector is positioned at around $-9^\circ$ of declination, the resolution is best around the same angle.

\section{Summary and outlook}
\label{sec7}
An analysis framework for reconstructing supernova neutrino burst directions at DUNE is described. By reconstructing the eES and $\nu_e$CC events during a supernova burst, the supernova direction can be determined from the correlation between the supernova neutrino and the outgoing primary electron created in the interaction with the LAr. The dominant interaction channels are $\nu_e$CC and eES, with only the latter containing significant accessible information on the neutrino direction, but being an order of magnitude less frequent than the former. 
Monte Carlo simulations in \larsoft{} were used to generate a data set of supernova bursts to evaluate the performance of the reconstruction algorithm for the planned DUNE detector setup. 
From this data set, the energy and direction of the primary electron were reconstructed. To reduce head-tail ambiguities a highly effective technique called brems flipping was developed.
Finally, the pointing resolution is derived for ensembles of supernova burst events using a maximum likelihood method. With perfect event classification of the two event types considered, the pointing resolution is 3.4\,degrees (6.6\,degrees) with 68\% coverage for supernovae at a distance of 10\,kpc and an effective fiducial volume of 40\,kton (10\,kton). For a moderately optimistic classification performance, incorrect classification of 4\% $\nu_e$CC events as eES, the estimated pointing resolution is 4.3\,degrees (8.7\,degrees) with 68\% coverage.  The results presented here represent an average over all supernova directions in the sky. 
In reality, there is a modest anisotropy in the burst reconstruction capability of the LAr TPC, resulting in a variation of the order of a few degrees in resolution. 

A continued effort will be carried out to further improve the current supernova pointing capabilities of DUNE. Prompt dissemination of directional information will be critical for multi-messenger astronomy, and efforts are underway to enable low-latency pointing with DUNE.  Furthermore, given a core collapse at the most frequently expected distance from Earth for a galactic supernova, detection of supernova neutrino events can be expected in multiple large-scale neutrino detectors. By combining DUNE's data with events in other detectors, one may take advantage of greater event statistics as well as the strengths of different detector technologies, resulting in a noticeable improvement to the current pointing resolution.  Furthermore, we anticipate that steady improvements to pattern recognition technologies, especially via machine-learning techniques, will enhance the performance of next-generation particle track reconstruction and head-tail disambiguation algorithms, allowing for additional gains in pointing ability. Extended investigation of the eES/$\nu_e$CC classification at DUNE is required, as this currently provides the greatest uncertainty in the pointing resolution.  Furthermore, subdominant interaction channels should be considered as well.  As a final note, laboratory investigation of neutrino interactions in the few-tens-of-MeV range in argon will be essential to fully understand the detector directional response~\cite{Asaadi:2022ojm}.

\section*{Acknowledgements}

The ProtoDUNE-XX detector was constructed and operated on the CERN Neutrino Platform. We gratefully acknowledge the support of the CERN management, and the CERN EP, BE, TE, EN and IT Departments for NP04/Proto-DUNE-SP. This document was prepared by DUNE collaboration using the resources of the Fermi National Accelerator Laboratory (Fermilab), a U.S. Department of Energy, Office of Science, Office of High Energy Physics HEP User Facility. Fermilab is managed by Fermi Forward Discovery Group, LLC, acting under Contract No. 89243024CSC000002. This work was supportedby CNPq, FAPERJ, FAPEG and FAPESP, Brazil; CFI, IPP and NSERC, Canada; CERN; MŠMT, Czech Republic; ERDF, Horizon Europe, MSCA and NextGenerationEU, European Union; CNRS/IN2P3 and CEA, France; INFN, Italy; FCT, Portugal; NRF, South Korea; Generalitat Valenciana, Junta de Andalucıa-FEDER, MICINN, and Xunta de Galicia, Spain; SERI and SNSF, Switzerland; TÜBİTAK, Turkey; The Royal Society and UKRI/STFC, United Kingdom; DOE and NSF, United States of America. This research used resources of the National Energy Research Scientific Computing Center (NERSC), a U.S. Department of Energy Office of Science User Facility operated under Contract No. DE-AC02-05CH11231.

\bibliography{mybib}

\appendix*
\section{Appendix}

The details of the code implementation of the descriped brems-flipping algorithm are shown in Algorithm~\ref{alg:df}.

\begin{algorithm}[H] 
\caption{Brems Flipping \label{alg:df}}
         \SetKwInOut{Input}{input}
        \SetKwInOut{Output}{output}
        \Input{Two candidate $e^-$ vertices with positions $\vec{r_1}$, $\vec{r_2}$ and momentum directions $\hat{d_1}$, $\hat{d_2}$;\\
        Set of secondary particle vertex positions $\{\vec{r_d}\}$}
        \Output{Reconstructed $e^-$ direction.}
    \SetKwBlock{Beginn}{beginn}{ende}
    \Begin{
        $SumCos_1$, $SumCos_2$ $\leftarrow 0$\;
        \For{$i \gets 1$ to $2$}{
            \For{$\vec{r} \in \{\vec{r_d}\}$}{
                $SumCos_i \leftarrow SumCos_i + \frac{\hat{d_i} \cdot (\vec{r} - \vec{r_i})}{\abs{\vec{r} - \vec{r_i}}}$ 
            }
        }
        
        \eIf{$SumCos_1$ $>$ $SumCos_2$}{
            \Return $\hat{d_1}$
        }{
            \Return $\hat{d_2}$
        }
    }
\end{algorithm}

\end{document}